\documentclass[useAMS,usenatbib]{mn2e}
\usepackage{aas_macros}
\usepackage[a4paper,centering, totalwidth=520pt, totalheight=700pt]{geometry}
\def\k#1 {k_{{\rm #1}}}

\def\mH2p{H_2^+}

\def\ltsima{$\; \buildrel < \over \sim \;$}
\def\simlt{\lower.5ex\hbox{\ltsima}}   
\def\gtsima{$\; \buildrel > \over \sim \;$}
\def\gtsim{\lower.5ex\hbox{\gtsima}}

\usepackage[fleqn]{amsmath} 
\usepackage{graphicx}
\usepackage{amssymb}
\usepackage{amsmath}
\usepackage{enumerate}
\usepackage{microtype}
\usepackage{color}
\usepackage{mathabx}

\title[RHAPSODY-G II]{RHAPSODY-G simulations II - Baryonic growth and metal enrichment in massive galaxy clusters }

\author[D. Martizzi et al.]{Davide Martizzi\thanks{Email:dav.martizzi@berkeley.edu}$^{1}$,
Oliver Hahn$^{2,3}$, 
Hao-Yi Wu$^{4,5}$, 
August E. Evrard$^{5}$ \newauthor 
Romain Teyssier$^{6}$ and 
Risa H. Wechsler$^{7,8}$ \\
\\
 $^{1}$ Department of Astronomy, University of California, Berkeley, CA 94720-3411, USA \\
 $^{2}$ Laboratoire Lagrange, Universit\'e C\^ote d'Azur, Observatoire de la C\^ote d'Azur, CNRS,\\
\quad Blvd de l'Observatoire, CS 34229, 06304 Nice cedex 4, France\\
 $^{3}$ Department of Physics, ETH Zurich, CH-8093 Z\"urich, Switzerland\\
 $^{4}$ California Institute of Technology, MC 367-17, Pasadena, CA 91125, USA\\
 $^{5}$ Departments of Physics and Astronomy, University of Michigan, Ann Arbor, MI 48109, USA\\  
 $^{6}$ Institute for Computational Science, University of Zurich, CH-8057 Z\"urich, Switzerland\\
 $^{7}$ KIPAC, Physics Department, Stanford University, Stanford, CA 94305, USA\\
 $^{8}$ SLAC National Accelerator Laboratory, Menlo Park, CA 94025, USA\\ 
 }

\begin{document}

\date{\today}
\pagerange{\pageref{firstpage}--\pageref{lastpage}} \pubyear{2014}
\maketitle

\label{firstpage}

\begin{abstract}
We study the evolution of the stellar component and the metallicity of both the intracluster medium and of stars in massive ($M_{\rm vir}\approx 6\times 10^{14}$ M$_{\odot}/h$) simulated galaxy clusters from the {\sc Rhapsody-G} 
suite in detail and compare them to observational results. The simulations were performed with the AMR code {\sc RAMSES} and include the effect of AGN 
feedback at the sub-grid level. AGN feedback is required to produce realistic galaxy and cluster properties and plays a role in mixing material in the central regions and regulating star formation in the central galaxy. 
In both our low and high resolution runs with fiducial stellar yields, we find that stellar and ICM metallicities are a factor of two lower than in observations. We find that cool core clusters exhibit steeper metallicity gradients 
than non-cool core clusters, in qualitative agreement with observations. 
We verify that the ICM metallicities measured in the simulation can be explained by a simple ``regulator'' model in which the 
metallicity is set by a balance of stellar yield and gas accretion. It is plausible that a combination of higher resolution and higher metal yield in AMR simulation would allow the metallicity of simulated clusters to match observed values; however this hypothesis needs to be tested with future simulations. Comparison to recent literature highlights that results concerning the metallicity of clusters 
and cluster galaxies might depend sensitively on the scheme chosen to solve the hydrodynamics. 
\end{abstract}


\maketitle

\section{Introduction}

Modelling the formation of realistic galaxy clusters in a cosmological context is one of the most challenging problems of galaxy formation. In 
the $\Lambda$CDM cosmological model, galaxy clusters are the most massive virialised structures and the latest objects to be assembled. 
Their formation depends hierarchically on the formation of their progenitors. 
While dark matter is the dominant component in the mass budget of galaxy clusters, the hot X-ray emitting intracluster medium 
(ICM) dominates the baryonic mass budget \citep[e.g.][]{2013ApJ...778...14G, 2013A&A...550A.131P}. The galaxies that populate the halo constitute only a tertiary 
component, with stars accounting for about 10 per cent of the baryonic mass (\citealt{2009ApJ...703..982G}; see e.g. our previous work \citealt{2015MNRAS.452.1982W}). Cluster centres are typically dominated by one (or a few) massive elliptical galaxies, the brightest cluster galaxies (BCGs). Deep optical and near 
infrared observations revealed the existence of the so-called intracluster light (ICL): extended haloes of diffuse light that typically surround BCGs and that are 
considered to be generated by the stellar material stripped from satellite galaxies \citep[e.g.][for recent studies]{2015A&A...579A.135L, 2015MNRAS.449.2353B}. 

The interaction of the various baryonic components with each other is complex and challenging to model. The ICM is stabilised against cooling by strong 
heating sources at the centre of clusters that are thought to be connected to active galactic nuclei via a phenomenon referred to as AGN feedback \citep{2006MNRAS.365...11C, 2007MNRAS.380..877S, 2012ARA&A..50..455F}. Such heating sources are supposed to 
regulate the supply of gas available in the most massive galaxies to form stars, either by preventing gas accretion and cooling or by triggering gas ejection events. 
The properties of the smallest satellite galaxies are influenced by stellar feedback that regulates star formation and may also eject metal-enriched 
gaseous material \citep{2010Natur.463..203G, 2013MNRAS.429.3068T, 2015arXiv150702282C}. Satellite galaxies are also stripped of their gaseous material by the ram pressure caused by their orbit through the ICM \citep{GunnGott:72} and part of 
their stellar and dark matter mass is prone to be stripped by tidal interactions with other galaxies and with the cluster potential \citep{1996Natur.379..613M, 2001ApJ...559..754M, 2006MNRAS.369.1021M, 2008MNRAS.383..593M, 2008MNRAS.387...79V}. As 
clusters evolve, they also accrete mass from the intergalactic medium (IGM) and merge with systems of similar size. The interplay between all these processes 
determines the budget of baryons in clusters. In particular, the cycle of accretion and ejection of gas from galaxies and from the cluster as a whole also 
determines the distribution of metals in the ICM and in stars. For this reason, studying the metallicity in the ICM and in cluster galaxies can provide very 
useful insights into the formation and evolution of clusters, as well as the physical processes at work. 

Because of their deep gravitational potentials, the most massive haloes retain the majority of their original baryon content 
\citep[e.g.][]{2013ApJ...778...14G, 2013A&A...550A.131P, 2014MNRAS.440.2077M}. Heavy elements produced by star 
formation in cluster galaxies should also be retained, and the $\sim 0.3$ solar metallicity observed in the ICM \citep{2014MNRAS.444.3581R} supports this expectation. 
However, attempts to employ chemical and population synthesis models fall short, by factors of several, of reproducing the metal content observed in Coma-scale clusters with $M \sim 10^{15}$ M$_{\odot}$: \cite{2014MNRAS.444.3581R} show that in order to simultaneously match the constraints on ICM metallicity and on the stellar mass to light ratios, the metal yield from supernovae has to increase with the cluster mass in a fashion apparently unjustified by chemical and population synthesis models; this fact lead \cite{2014MNRAS.444.3581R} to conclude that currently available data on massive clusters may be faulty. Such a {\itshape conundrum} provides a strong motivation for performing a detailed study of the metallicity content of massive clusters via numerical simulations.

The history of theoretical studies of galaxy clusters is long and one of the approaches that has been followed by several authors to model these systems and to understand the origin of their metal content is through semi-analytical models coupled to N-body simulations \citep{2008MNRAS.386...96C, 2009A&A...504..719K,  2010ApJ...716..918A, 2013MNRAS.428.1225S} or cosmological hydrodynamical simulations \citep{2003MNRAS.339.1117V, 2006MNRAS.371..548R, 2007MNRAS.382.1050T, 2008MNRAS.391..110D, 2011MNRAS.415..353W, 2013MNRAS.432.3005C, 2013ApJ...763...38S}. Most state-of-the-art cosmological zoom-in simulations of galaxy clusters reach a spatial resolution of $\sim$kpc 
\citep{2008ApJ...687L..53P, Teyssier:2011, 2012MNRAS.420.2058E, Martizzi2014b, 2014MNRAS.441.1270L, 2015ApJ...806...68L, 2015arXiv150306065S}. Such 
simulations are challenging since they require a very large dynamic range of scales to be captured: from the scales at which galaxies assemble their stars to the 
scale at which cosmological gas accretion onto clusters happens. With the maximum spatial resolution of $\sim$kpc achievable in current state-of-the-art simulations, 
phenomena like star formation, gas 
accretion onto the supermassive black holes (SMBHs) at the centre of galaxies and their feedback onto the surrounding gas are beyond direct numerical reach. 
Instead, they have to be treated at sub-resolution 
scales using phenomenological models motivated by simple physical assumptions tied to the resolved scales. Such simulations have been shown to be more successful at 
reproducing some of the properties of observed clusters when AGN feedback is included \citep{Sijacki:2006, 2009MNRAS.398...53B, Teyssier:2011, Martizzi2012a, 
Martizzi2014b, 2014MNRAS.441.1270L}. However, as already mentioned, the methods and models adopted to simulate clusters and unresolved phenomena are highly 
simplified compared to reality, and careful comparison to available observational data might shed light on whether their validity breaks down in certain 
regimes. With this goal in mind, in this paper we consider cosmological hydrodynamical simulations of the most massive galaxy clusters in the Universe 
($M_{\rm vir}\sim 10^{15}$ M$_{\odot}$). The simulations we analyse belong to the {\sc Rhapsody-G} suite that is described in \cite{2015arXiv150904289H} (paper~I) and 
that has been already used by \cite{2015MNRAS.452.1982W} to study the baryon content of clusters and the properties of baryonic scaling relations. As shown in paper~I, the {\sc Rhapsody-G} simulations reproduce a variety of cluster observables including mass profiles, SZ scaling relations and the existence of two thermodynamically distinct populations of clusters (cool core and non-cool core). 

The goal of this paper is to study the distribution of metallicity in galaxy clusters and highlight the phenomena that determine it. This comparison will 
identify the strengths and limits of the current simulation techniques. We will focus on simulations performed with adaptive mesh refinement (AMR) methods. \cite{2014MNRAS.438..195P} recently published a similar analysis based on 
cosmological smoothed particle hydrodynamics (SPH) simulations. In fact, unlike AMR codes (and grid codes in general), 
standard SPH does not allow exchange of metals between resolution elements, unless it is explicitly implemented. 
Such differences between the two most common implementations of hydrodynamics are worth to be studied in detail and can be extremely important for setting the stellar mass-metallicity relation in simulated galaxies \citep{2015arXiv150708281S} and the metallicity distribution in general. Comparing our results to those discussed in the literature constitutes another important goal for our paper, so that poorly understood numerical and modelling uncertainties present in current simulations of galaxy clusters can be identified. {Most of the recent studies of cluster metallicities in cosmological hydrodynamical simulations with AGN feedback have been performed with SPH methods and older studies did not include AGN feedback \citep[e.g.][]{2007MNRAS.382.1050T}, a fact that constitutes the main motivation of our work. }

The paper is structured as follows. Section 2 describes the methods and models adopted for the {\sc Rhapsody-G} simulations. Section 3 is dedicated to the stellar properties, metallicities in particular, and we discuss caveats and 
effects related to the limited resolution of the simulations. Section 4 discusses the results we obtain for the ICM metallicity. In Section 5, we provide a 
simple analytical model to interpret the results. Section 6 compares our simulations to recent work by other groups, with the goal of capturing a snapshot of 
the current state of cosmological simulations of clusters. Finally, in Section 7 we summarise our results and discuss the future perspective of 
the {\sc Rhapsody-G} project.

\section{The Simulations}

We use the {\sc Rhapsody-G} set of ten galaxy clusters in this study, which is a carefully selected subset of clusters of the same virial mass at $z=0$ from a large cosmological volume. The clusters were chosen to capture the full range of accretion histories leading to the same final mass in order to quantify the effect of cosmic variance on cluster properties. 
In this section, we briefly summarise the technical details and the methods used for the simulations.  For a detailed description 
of the simulation suite and of the physical models we adopted, we kindly refer the reader to paper I \citep{2015arXiv150904289H}. 

\subsection{The {\sc Rhapsody} sample of galaxy clusters}
We analyse hydrodynamical zoom-in simulations of 10 galaxy clusters from the original {\sc Rhapsody} set of 96 haloes 
\citep[][there studied using $N$-body simulations]{Wu2013a, Wu2013b} of mass $M_{\rm vir}=10^{14.80\pm 0.05} h^{-1}$ M$_{\odot}$. 
These haloes were identified at redshift $z=0$ in a cosmological box of volume 
$1 h^{-3}$ Gpc$^3$ from the {\sc LasDamas} simulations suite. 

We show the results of re-simulations of zoom-in regions \citep[set-up using {\sc Music}][]{2011MNRAS.415.2101H} centred on ten different clusters: nine of the central haloes are chosen to have a mass 
$M_{\rm vir}\approx 6\times 10^{14}$ M$_{\odot}/h$ and the tenth has $M_{\rm vir}\approx 1.3 \times 10^{15}$ M$_{\odot}/h$. Three of the main haloes have very high 
concentration, two have an extreme number of subhaloes, and five have approximately the median concentration and typical number of subhaloes. With reference to the 
original {\sc Rhapsody} sample, the ID numbers of the haloes we re-simulate are 211, 337, 348, 361, 377, 448, 474, 545, 572, 653.

\subsection{Cosmology, numerical methods and resolution}
The {\sc Rhapsody-G} simulations we analyse in this paper were performed with the AMR code {\sc Ramses} \citep{2002A&A...385..337T}.  We simulate cosmic structure formation in a flat universe with 
a cosmological constant, cold dark matter and baryons, the standard $\Lambda$CDM scenario. We choose cosmological parameters as in the original {\sc Rhapsody} suite, i. e. $\Omega_{\rm M}=0.25$, 
$\Omega_{\rm \Lambda}=0.75$, $\Omega_{\rm b}=0.045$, $h=0.7$. Note that the cosmic baryon fraction assumed in these simulations 
($\Omega_{\rm b}/\Omega_{\rm M}=0.18$) is thus slightly higher than what has been found by the \cite{2015arXiv150201589P} and this may lead our simulations to underestimate gas metallicities for a fixed amount of star-formation and hence metal production. 

The computational domain is a cubic box of side 1 $h^{-1}$ Gpc. The {\sc Rhapsody-G} were performed at two different resolutions. The lower resolution simulations are labeled as ``4K''. In these runs, we chose 
an initial level of refinement $l=12$ ($4096^3$), but allowed to dynamically refine down to level $l=18$. This choice corresponds to a dark matter particle mass 
$m_{\rm dm}=8.22 \times 10^{8} h^{-1}$ M$_{\odot}$ and an initial baryonic matter resolution element 
$m_{\rm b}= m_{\rm dm} \Omega_{\rm b}/(\Omega_{\rm M} - \Omega_{\rm b})=1.80 \times 10^{8} h^{-1}$ M$_{\odot}$. The minimum allowed mass for the stellar 
particles is $m_{\rm *,min}=0.2\times m_{\rm b}=3.6 \times 10^{7} h^{-1}$ M$_{\odot}$. The maximum spatial (minimum cell size) resolution reached in these simulations is $\Delta x = 3.8 h^{-1}$ kpc (in physical units). The cell size also provides 
the value of the gravitational softening, since the Poisson equation is solved on the AMR grid. All the R4K simulations have been evolved down to redshift $z=0$.

Haloes 545, 572 and 653 have also been simulated at higher resolution. These runs are much more expensive and we could only evolve halo 653 down to $z=0$, whereas halo 545 and 572 have only 
been evolved down to redshift $z=0.5$. We label these simulations as ``8K''. The mass resolution achieved in these runs is 8 times better than in the R4K case, i.e. 
$m_{\rm dm}=1.03 \times 10^{8} h^{-1}$ M$_{\odot}$, $m_{\rm b}= 2.25\times 10^{7} h^{-1}$ M$_{\odot}$ and $m_{\rm *,min}=4.5\times 10^{6} h^{-1}$ M$_{\odot}$. 
The minimum cell size achieved in the R8K runs is $\Delta x = 1.9 h^{-1}$ kpc (in physical units). 

\begin{table*}
\begin{center}
{\bfseries Summary of the current {\sc Rhapsody-G} simulations}
\end{center}
\begin{center}
\begin{tabular}{lccccc}
\hline
 Resolution & $\Delta x$ [kpc$h^{-1}$]& $m_{\rm dm}$ [M$_{\odot}h^{-1}$]& $m_{\rm b}$ [M$_{\odot}h^{-1}$] & $m_{\rm *,min}$ [M$_{\odot}h^{-1}$] & Final redshift \\
\hline
R4K & 3.8 & $8.22 \times 10^{8}$ & $1.80\times 10^{8}$ & $3.6 \times 10^{7}$ & $z=0.0$ \\
R8K & 1.9 & $1.03 \times 10^{8}$ & $2.25\times 10^{7}$ & $4.5 \times 10^{6}$ & $z=0.5$ \\
\hline
\end{tabular}
\end{center}
\caption{Parameters of all the simulations used in this paper. Col. 1 resolution label. Col. 2: cell size (physical). Col. 3: smallest dark matter particle mass. 
Col. 4: baryonic mass element. Col. 5: smallest stellar particle mass. Col. 6: final redshift of the simulation. Notice that only halo 653 has been evolved 
down to redshift $z=0$ in the R8K version.}
\label{tab:parameters}
\end{table*}

The parameters of all simulations are summarised in Table~\ref{tab:parameters}.

\subsection{Subgrid modelling and baryonic processes}

The implementation of the physics of baryons in the {\sc Rhapsody-G} simulations is limited by the spatial and mass resolution we achieve. We resort to sub-grid modelling of many processes 
that happen at scales much smaller than our cell size, but that are relevant to study galaxy formation and the properties of galaxy clusters. 

Gas radiative cooling is implemented using rates based on \cite{1993ApJS...88..253S} and H, He and metal line cooling are taken into account. Gas metallicity is advected with the flow as a passive scalar 
and is taken into account in the cooling function in a self-consistent way. The solar mixture of \cite{1989GeCoA..53..197A} is assumed to compare our results to observational data. We also consider the effect 
of the UV background according to \cite{1996ApJ...461...20H}, but we set the epoch of reionisation 
to redshift $z_{\rm reion}=10$; this choice has been made to take into account early reionisation in the highly biased proto-cluster regions that we simulate. 

As a rough model for unresolved thermal and turbulent motions in the ISM, we introduce a temperature floor for the high density gas that cools to low 
temperatures and ends up contributing to the ISM of galaxies:
\begin{equation}
T_{\rm floor}=T_*\left( \frac{n_{\rm H}}{n_*} \right)^{\Gamma -1}
\end{equation}
where $n_*=0.1$ cm$^{-3}$ is the threshold defining star-forming gas and $T_*=10^4$ K is a characteristic temperature mimicking thermal and turbulent motions. The polytropic exponent 
$\Gamma=5/3$ controls the stiffness of the equation of state. In practice, gas cannot cool below the temperature floor, but can be heated above it.

Star formation is implemented via a simple sub-grid model. We create new star particles in cells with gas density larger than $n_*$. The mass of the star particles depends on resolution and in our simulations 
is set to $m_{\rm star}=0.2 \times m_{\rm b}$, where $m_{\rm b}$ is the baryonic matter resolution element. The local star formation rate is set by:
\begin{equation}
\dot{\rho}_*=\epsilon_*\frac{\rho_{\rm gas}}{t_{\rm ff}}
\end{equation}
where $t_{\rm ff}=(3\pi/32 G \rho)^{1/2}$ is the free-fall time $\epsilon_*$ is the star formation efficiency per free-fall time. In our fiducial runs we set $\epsilon_*=0.02$. Stellar particles are spawned locally following 
a Poisson process. At formation time, each stellar particle is assigned a metallicity equal to the metallicity of the gas in the cell where the particle is formed.

We use the standard (thermal) supernova feedback implementation in the {\sc Ramses} code \citep{2008A&A...477...79D}. Every time a star particle is formed, we assume that a fraction $\eta = 0.1$ is ejected by 
supernovae (SNe) after 10~Myr. The total energy per supernova is $10^{51}$ erg. We also assume that SNe produce 1~M$_{\odot}$ of metals per 10~M$_{\odot}$ of ejecta; this corresponds to a metal yield as a 
function of ejecta mass of $y=0.1$. In other words, a fraction $y\eta = 0.01$ of every star particle formed is assumed to be converted into metals during stellar evolution and is then ejected into the ISM. For each SN event we add the mass of the ejecta to the gas density field and update the local value of the passive scalar that traces the gas metallicity and is advected with the gas flow. Even if slightly more conservative than typical values $y\eta\sim 0.02$, the value for the metal yield has been chosen to match, on average, the metal enrichment from massive stars as computed by \cite{1995ApJS..101..181W} and the values used in previous simulations of galaxy clusters performed with the {\sc ramses} code \citep{Teyssier:2011, Martizzi2014b}. {This model does not explicitly include the difference between SNe type II and SNe type Ia which happen on different time scales and have different metal yields; given the time scale adopted for the delay of SN event and the chosen metal yield, the SN feedback scheme we adopt is appropriate only for SN type II.} {\itshape In Appendix~\ref{appendix:b} we show the effects of assuming a higher metal yield on the simulated galaxy population; these results are very relevant for our future work.}

Our simulations also include a sub-grid model for active galactic nucleus (AGN) feedback inspired by the thermal feedback models of \cite{2005MNRAS.361..776S} and \cite{2009MNRAS.398...53B}. This 
model has been previously used by some of the authors and was shown to successfully prevent gas overcooling in galaxy clusters of typical mass ($M_{\rm vir}\sim 10^{13.5-14.5}$ M$_{\odot}$) \citep{Teyssier:2011, Martizzi2012a, Martizzi2012b, Martizzi2014a, Martizzi2014b}. 

\subsection{AGN feedback sub-grid model}

The AGN feedback scheme in {\sc RAMSES} has been modified compared to the original implementation by \cite{Teyssier:2011}; in this section we discuss its new features. 

According to current theory, primordial massive black holes can be formed either as the end-product of the collapse of Pop III stars \citep{2001ApJ...551L..27M}, or as result of 
direct collapse of baryonic material within low angular momentum haloes \citep{2003ApJ...596...34B, 2006MNRAS.370..289B}. Given these considerations, it is natural to associate sink formation in 
cosmological simulations only to gas properties. Our simulations are the first cosmological simulations of galaxy clusters that use the algorithm developed by \cite{2014arXiv1412.0510B} to 
identify gravitationally bound gaseous structures on-the-fly. Supermassive black holes (SMBHs) form within the gas clumps and are modelled as sink particles. 
Gas clumps are detected when contiguous regions of high density gas exceed $10^{-29}\,{\rm g/cm^3}$ (comoving units).
Sink particles are formed within clumps only if the following conditions are met:
\begin{itemize}
  \item The clump does not contain a sink particle within its boundaries.
  \item The clump is gravitationally bound.
  \item The accretion rate onto the central regions of the clump is high enough.
\end{itemize}
The accretion rate used for the third condition is computed as
\begin{equation}
\dot{M}_{\rm clump}=\frac{M_4}{t_{\rm ff}}
\end{equation}
where $t_{\rm ff}$ is the local free-fall time and $M_4$ is the gas mass enclosed within a spherical region of radius equal to $R_4=4\Delta x$, where $\Delta x$ is the cell size. 
To allow sink formation only in the most massive high-redshift haloes we form a sink only if $\dot{M}_{\rm clump}>30$~M$_{\odot}$/yr. 

The trajectory of sink particles is integrated as if they were N-body particles. We also include a sub-grid model for the drag force experienced by a black hole from the gas in the wake it forms as it moves \citep{1999ApJ...513..252O}. We merge two sinks only if the kinetic energy associated associated to their relative motion is lower than the potential energy of the two body system. 

By assuming that the SMBH initially accretes mass at the Eddington rate and that the SMBH is able to heat the surrounding gas to a temperature of $10^7$~K during the Salpeter time $t_{\rm S}$, we get an estimate for the mass of the seed SMBH:
\begin{equation}
M_{\rm BH,s}=\frac{10^{-5}}{\epsilon_{\rm c}}\dot{M}_{\rm clump}t_{\rm S}.
\end{equation}

Each SMBH accretes mass at a rate that depends on the conditions of the flow around the sink. In the original version of the scheme SMBHs were accreting at a rate proportional to the Bondi--Hoyle accretion rate times a density-dependent boost factor. In quasi-spherical flows the Bondi-Hoyle accretion rate formula can overestimate the accretion rate in case of cold, supersonic gas accretion \citep{2012MNRAS.421.3443H}. In this case the accretion rate is better approximated by the free-fall rate:
\begin{equation}\label{mdot_ff}
  \dot{M}_{\rm ff}=\frac{M_{\rm gas}(r<\lambda_J)}{t_{\rm ff}(r<\lambda_J)}
\end{equation}
where $M_{\rm gas}(r<\lambda_J)$ is the gaseous mass enclosed within a sphere of radius equal to the Jeans length $\lambda_J$, and $t_{\rm ff}(r<\lambda_J)$ is the free-fall time in the same region. In our scheme, we interpolate between the Bondi-Hoyle regime and the free-fall regime. We compute the accretion rate onto SMBHs as:
\begin{equation}
\dot{M}_{\rm BH}=4\pi\alpha_{\rm boost}\tilde{r}_B^2v_B\rho
\end{equation}
where $\alpha_{\rm boost}$ is a density-dependent factor that accounts for unresolved multiphase turbulence in the SMBH environment \citep{2009MNRAS.398...53B}, 
$v_B=\sqrt{u^2+c_{\rm s}^2}$, and $\tilde{r}_B$ is a modified Bondi radius defined by 
\begin{equation}\label{modified_bondi_radius}
\tilde{r}_B=\min(r_B,4\Delta x)
\end{equation}
with $r_B$ equal to the standard Bondi radius:
\begin{equation}
r_B=\frac{G M_{\rm BH}}{v_B^2}.
\end{equation}
The density-dependent boost factor $\alpha_{\rm boost}$ is defined by:
\begin{eqnarray}
\nonumber 
\alpha_{\rm boost}&=&\left( \frac{n_{\rm H}}{n_*} \right)^2~~{\rm if }~~n_{\rm H} > n_* = 0.1~{\rm H/cc,}\\
\alpha_{\rm boost}&=&1~~~{\rm otherwise.}
\end{eqnarray}
The choice for this particular form of $\alpha_{\rm boost}$ is strictly dependent on the chosen equation of state for the ISM. In eq.~(\ref{modified_bondi_radius}), we limit the modified Bondi radius 
to a maximum value of four times the cell size $\Delta x$, i.e. to the minimum resolved Jeans length. With this choice, in the case of cold, supersonic ($u\gg c_{\rm S}$) gas accretion we 
recover the free-fall rate in eq.~(\ref{mdot_ff}). In the case of hot gas accretion, the formula simply reduces to the Bondi-Hoyle formula used in our older scheme.

SMBHs are not allowed to accrete at a rate that exceeds the Eddington limit 
\begin{equation}
\label{eddington_formula}
\dot{M}_{\rm ED}=\frac{4\pi{\rm G}M_{\rm BH} m_{\rm p}}{\epsilon_{\rm r} \sigma_{\rm T}c}~~{\rm with}~~\epsilon_{\rm r} \simeq 0.1{\rm .}
\end{equation}
where $\epsilon_{\rm r}$ is is the efficiency at which accreting gas rest mass energy is converted into radiation. To enforce this upper limit we always set the accretion rate to
\begin{equation}
\label{accretion_formula}
\dot{M}_{\rm acc}=\min (\dot{M}_{\rm BH},  \dot{M}_{\rm ED})
\end{equation}
At each time step, a total gas mass of $\dot{M}_{\rm acc} \Delta t$ is removed from all cells within the sink radius. In order to prevent the gas density from vanishing or becoming negative, we do not remove more than 50 per cent of the gas at each time step.

At each time step we compute the thermal energy injected in the gas surrounding each black hole as
\begin{equation}
\Delta E = \epsilon_{\rm c} \epsilon_{\rm r}\dot{M}_{\rm acc}c^2 \Delta t {\rm .}
\label{enerdump}
\end{equation}
where $\epsilon_{\rm c}$ is the coupling efficiency, i.e. the fraction of radiated energy that is coupled with the surrounding gas. The correct value for $\epsilon_c$ can be set by requiring the simulations 
to reproduce the observed $M_{{\rm BH}}-\sigma$ relation; we use the fiducial value $\epsilon_{\rm c} \simeq 0.15$ \citep{2009MNRAS.398...53B}. The energy $\Delta E$ is not immediately injected 
in the gas, but is accumulated and stored in a new variable $E_{\rm AGN}$, so that we can prevent the gas from instantly radiating away this energy via atomic line cooling. We release this energy within the 
sink radius when 
\begin{equation}\label{eq:min_ene}
E_{\rm AGN} > \frac{3}{2} m_{\rm gas} k_{\rm B} T_{\rm min},
\end{equation}
where $m_{\rm gas}$ is the gas mass within the sink radius and $T_{\rm min}$ is the minimum feedback temperature. We distribute the energy within the sink radius (a sphere of radius 4 cells) in a mass-weighted fashion. 
$T_{\rm min}$ should be chosen to be at least $10^7$ K, the temperature above which line cooling is not very efficient. In our simulations we adopt the fiducial value $T_{\rm min}=10^7$ K, the same value used in previous {\sc Ramses} simulations \citep{Teyssier:2011}. We have performed detailed testing of the robustness of the model against parameter variation in \cite{2015arXiv150904289H}. 

We would like to stress that ongoing improvements to the AGN feedback scheme are currently being implemented in the {\sc Ramses} code. The description and the calibration of the improved model will be described in detail by Biernacki et al. (in prep.).

\subsection{Identification of haloes and analysis procedure}

We used the phase-space halo finder {\sc Rockstar} \citep{2013ApJ...762..109B} to identify sub-haloes and to measure their position and properties. We use the dark matter density peak as the centre of the halo, which closely coincides with the stellar mass density peak in most cases. 

Our analysis in this paper is focused on the main halo hosting the cluster. The main halo is identified using the {\sc Rockstar} at the lowest redshift available ($z=0$ for the R4K runs and $z=0.5$ for the R8K runs) and 
tracked back in time by taking the most massive progenitor. The latter is defined as the most massive halo whose particle distribution overlaps with the present halo. Once the centres of the most massive progenitors of the main halo are known, their properties are extracted and analysed. 

In this paper we will focus on the dark matter, stellar and gas properties in the context of a model aimed at explaining the metallicity of the intracluster medium. Since we are interested in studying the average profiles of the simulated clusters and since they all belong to a very narrow mass bin, we compute the stacked profiles of all the relevant quantities. We stack data from the R4K and the R8K runs separately, because this will allow us to assess convergence as a function of numerical resolution. Hence, where it is not explicitly stated, we consider stacked properties. 

\section{Evolution of the stellar content}

In this section, we analyse the evolution of the stellar content of the galaxy clusters in the {\sc Rhapsody-G} sample with particular emphasis on the properties of the stellar population and its metallicity. We also show the difference between the stellar properties in cool core and non-cool core clusters. 

\subsection{Growth of the stellar mass profile and star formation rates}
\begin{figure*}
\begin{center}
\includegraphics[width=0.99\textwidth]{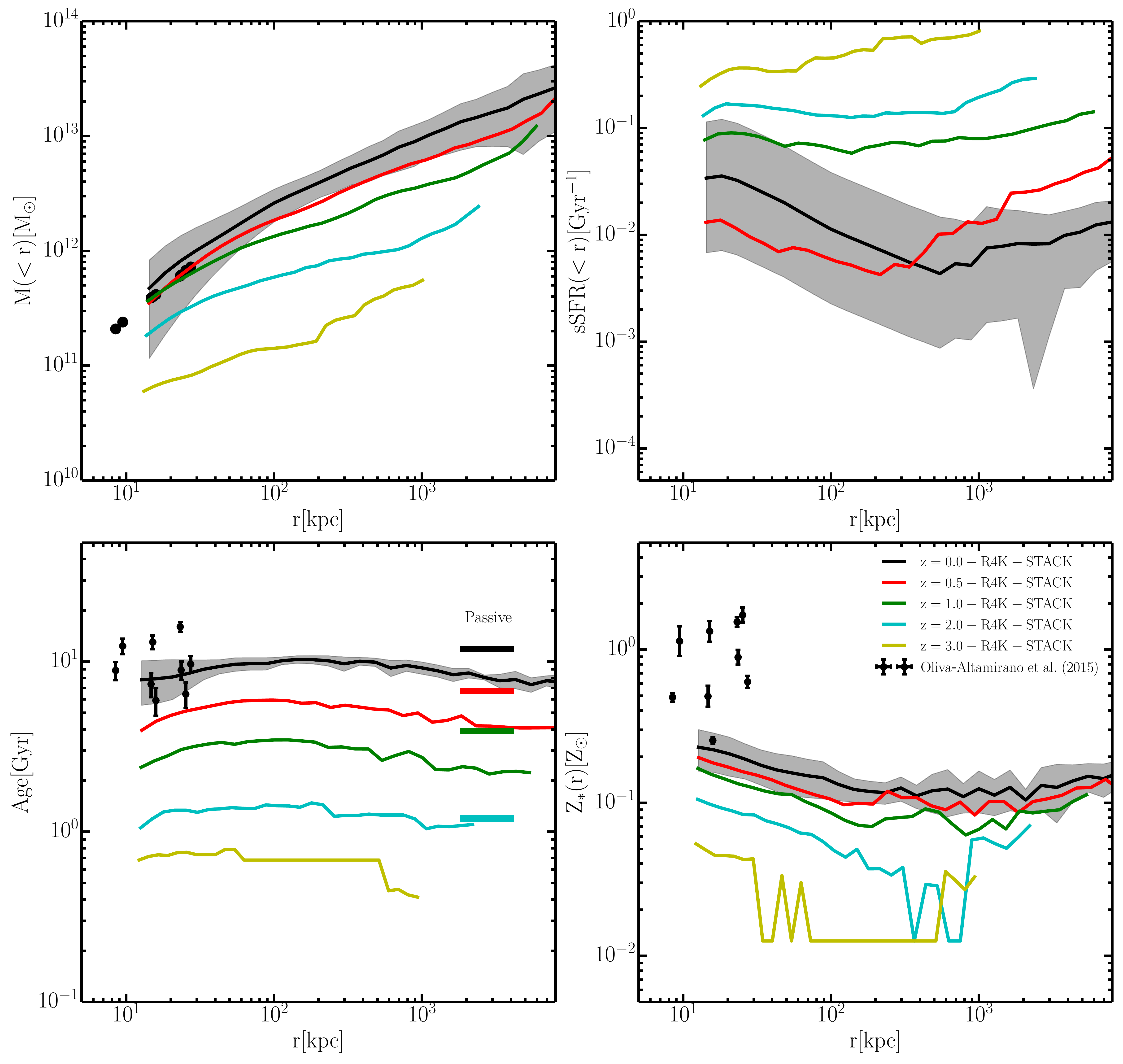}
\end{center}
\caption{\label{fig:stprop4k}Evolution of the stellar properties for the R4K runs (solid coloured lines). The minimum radius plotted for all the profiles 
is $r_{\rm min}=2\Delta x$, and the maximum radius is $r_{\rm max}\sim4-6R_{\rm 200,m}$ at high redshift and 8 Mpc at $z\leq 0.5$. Top left: cumulative stellar mass. Top right: specific star formation 
within a given radius averaged over a time interval $\Delta t=10^8$ yr. Bottom left: mean stellar age. The thick coloured lines represent a passively evolving 
stellar population of age 1.2~Gyr at redshift $z=2$. 
Bottom right: stellar metallicity. In all panels the dark shaded areas 
represent the typical 1-$\sigma$ scatter among different haloes at redshift $z=0$. As in Figures 1 and 2, the observational data from the sample of BCGs at 
redshift $z<0.095$ of 
Oliva-Altamirano et al. (2015) is plotted (black points with error bars); their data is extrapolated to and plotted at twice the effective radius of the 
galaxy; the extrapolation is done using the gradients measured and reported in their paper. }
\end{figure*}

\begin{figure*}
\begin{center}
\includegraphics[width=0.99\textwidth]{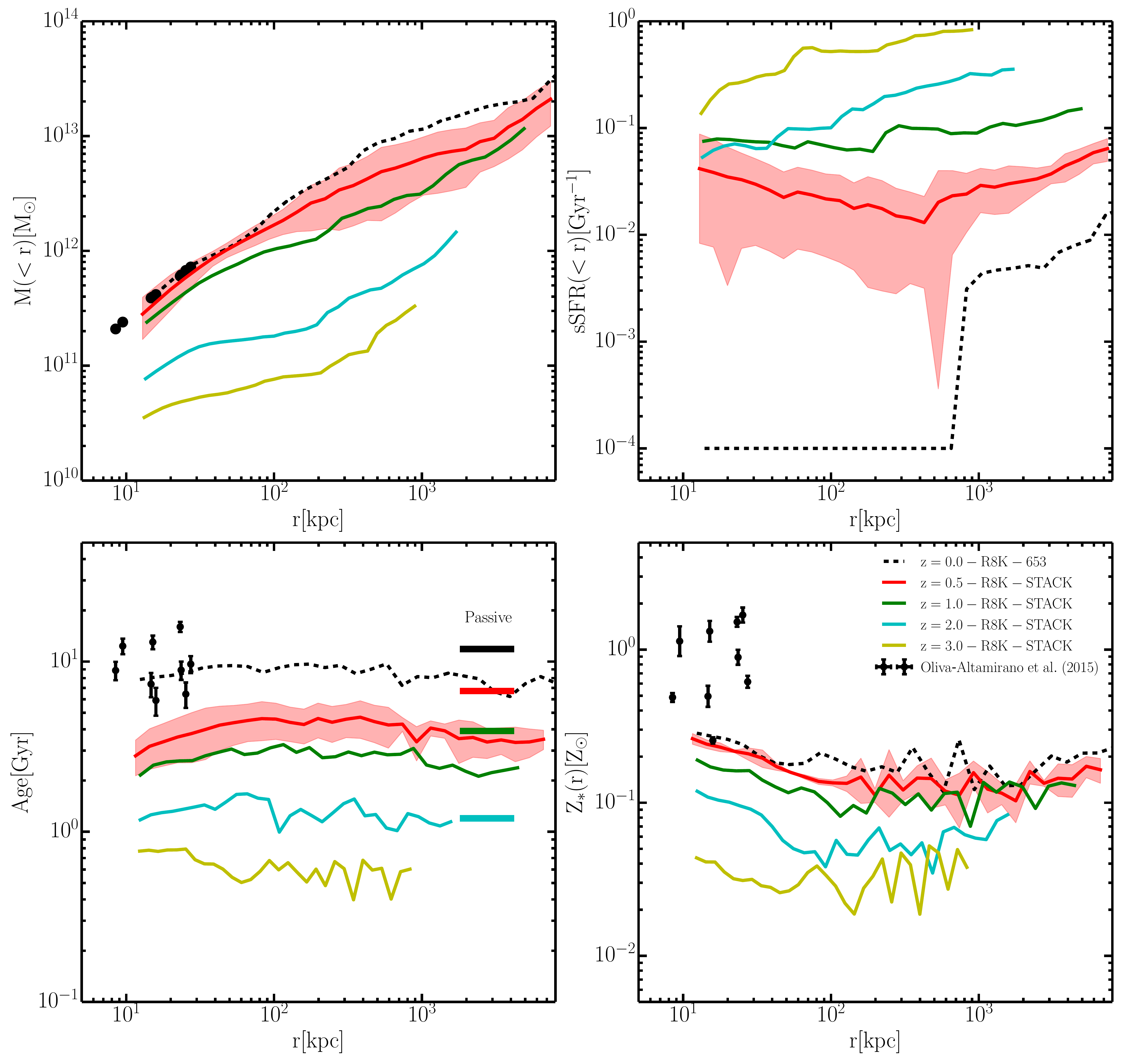}
\end{center}
\caption{\label{fig:stprop8k} Same as Figure~\ref{fig:stprop4k}, but for R8K. The redshift $z=0$ line is available only for halo 653 (black dashed line). The red shaded areas 
represent the typical 1$\sigma$ scatter between different haloes at redshift $z=0.5$.}
\end{figure*}

We first discuss the evolution of the stellar properties of the cluster as a function of redshift as summarised by Figures~\ref{fig:stprop4k} 
(4K runs) and \ref{fig:stprop8k} (8K runs). The results at redshift $z=0$ for halo 653 in its R8K versions are also shown but since they are only 
for one halo, we cannot extend our conclusions to the whole sample of clusters at redshift $z=0$ and R8K resolution. The top left panel of 
the figures shows the evolution of the stellar mass profile as a function of redshift. 
Within the central 50~kpc of the cluster where the brightest cluster galaxy (BCG) sits, the mass profiles shows a very rapid evolution at high redshift 
$z>1$, whereas it does not significantly evolve at lower redshift. This fact indicates that most of the mass in the BCG is assembled at high redshift but its growth  continues at a slower rate at $z<1$, over which time the BCG grows only by a factor $\lesssim 2$. The evolution of the mass of BCGs has been extensively 
studied in recent years in both simulations 
\citep{2007ApJ...668..826C, 2009ApJ...696.1094R, 2013ApJ...771...61L, 2013MNRAS.435..901L} and observations 
\citep{2002MNRAS.329L..53B, 2002ApJ...566..103N, 2002ApJ...567..144N, 2008MNRAS.387.1253W, 2009MNRAS.395.1491B, 2012ApJ...759...43T, 
2013MNRAS.433..825L, 2013MNRAS.434.2856B, 2014MNRAS.440..762O, 2015MNRAS.449.2353B}. The common picture that emerged is that most of the BCG mass is assembled by 
$z\sim 2$ and its subsequent mass evolution is largely determined by dry mergers that are responsible for a mass growth of a factor $\lesssim 2$ 
between redshifts $z \sim 1$ and $z \sim 0$. \cite{2015MNRAS.449.3347O} report effective radii and dynamical masses of 9 BCGs observed at redshift 
$z<0.095$ with VIMOS in IFU mode. We over-plot the dynamical masses of these 9 BCGs in Figures~\ref{fig:stprop4k} (4K runs) and \ref{fig:stprop8k} as 
black points; to better compare to our profiles, we assume that the dynamical mass is reached at a distance of twice the BCG effective radius. Based on these assumptions, we conclude that our simulations at $z=0$ agree with the observational data reasonably well. However, the {\sc Rhapsody-G} simulations 
presented here only have limited resolution: the stellar mass profiles appear not to have fully converged, especially at the highest redshifts. 
Simulations with even higher resolution than 8K are needed to properly check convergence. We will expand our discussion of this issue in the following sections.

\subsection{Specific star formation rates}

The top right panels of Figures~\ref{fig:stprop4k} (4K runs) and \ref{fig:stprop8k} (8K runs) show the cumulative specific star formation rates sSFR$(<r)$ in the cluster as a function of radius, i.e. the average sSFR within a sphere of given radius $r$. The sSFR is measured averaging over a time interval $\Delta t=10^8$ yr. This figure allows us to determine the radius within which star formation is effectively quenched. The evolution of the sSFR profiles shows that star formation at low redshift $z<1$ is a factor 5 to 10 lower than the star formation rate at higher redshift at all radii. This decrease of the specific star formation rates is a combined effect of the suppression of star formation in the satellite galaxies due to environmental effects in the cluster environment (ram pressure stripping, etc.), but is mostly driven by AGN feedback in the centrals and BCGs that possess a SMBH \citep{Martizzi2012a, 2013MNRAS.436.1750R, Martizzi2014b}. In clusters with lower mass than those in the {\sc Rhapsody-G} sample, thermal AGN feedback is usually extremely efficient at suppressing star formation in the central regions of the cluster.  As can be appreciated by Figure~\ref{fig:stprop8k}, star formation in the R8K version of halo 653 is almost totally suppressed for $r<700$~kpc at redshift $z=0$ (the sSFR saturates to the minimum threshold we allow in the plot; the star formation rate is numerically 0). However, when looking at the stacked profiles we find an increase of the sSFR towards the centre for $r\lesssim 500$~kpc at redshift $z<0.5$ --- even if the sSFR values are still very low. The fact that the sSFR increases towards the centre is related to the presence of 4 cool core clusters in the sample (haloes 361, 377, 448 and 545) which have some residual star formation in the centre. If the cool core and non-cool core clusters are split into two separate samples a huge difference is seen in the sSFR for $r\lesssim 500 $ kpc at redshift $z=0$ (see Subsection~\ref{sec:cc_ncc}). Recent observational results \citep[e.g.][]{2012MNRAS.423..422L} suggest that star-forming BCGs are associated to cool core clusters, in agreement with our results. It is worth noting that at large radii $r\gtrsim 500$ kpc at redshift $z<0.5$, the sSFR starts increasing again, suggesting that infalling satellite galaxies are still able to continue part of their star formation activity during their way through the outskirts of the clusters, but overall experience a significant decrease in sSFR as they approach the cluster. We will explore the properties of satellite galaxies, in particular the quenched fraction and their kinematics in a future paper.

\subsection{Mean age of the stellar population}

The bottom left panels of Figures~\ref{fig:stprop4k} (4K runs) and \ref{fig:stprop8k} (8K runs) show the mass-weighted average age of the stellar population 
as a function of radius. We find that the mean stellar age is very homogeneous as a function of radius. Significant evolution is only seen as a function of redshift. 
This fact suggests that most of the stellar mass is formed early on and simply ages as the cluster grows. The thick horizontal lines on the right side of the bottom left panels of 
Figures~\ref{fig:stprop4k} and \ref{fig:stprop8k} show the age of a passively evolving stellar population of age 1.2 Gyr at redshift $z=2$. The typical stellar age in the simulations is lower than that of a passively evolving stellar population because of ongoing star formation producing new stars. However, newly formed stars never dominate the mass budget and the mean stellar age in 
the simulated clusters never differs from the age of the passively evolving population by more than a factor $\sim 2$. The largest difference between a pure passively evolving stellar population and the simulations is observed at $z=0.5$. It is worth noticing that the central stellar 
ages at $z=0$ match the ages of the stellar populations in the BCGs observed by \cite{2015MNRAS.449.3347O} (black points). Note that 
the \cite{2015MNRAS.449.3347O} data points have been extrapolated to a distance of twice the BCG effective radii using the gradients reported in their paper. 

\subsection{Stellar metallicity}

The bottom right panels of Figures~\ref{fig:stprop4k} (4K runs) and \ref{fig:stprop8k} (8K runs) show the mass-weighted stellar metallicity averaged
in spherical shells at radius $r$. One important constraint coming from \cite{2015MNRAS.449.3347O} (black points) is that the stellar metallicity of the BCGs 
is very close to the solar value and even super-solar for some BCGs. The central stellar metallicity found in our simulations is close to the lower envelope 
formed by the observed BCGs. The stacked profile we show is only an average, but none of the individual BCGs in our sample has super-solar metallicity. This 
fact might indicate that a physical element necessary for BCG metallicities to reach values close to solar is missing in our simulations. Given the evolution of the 
stellar age, a possible solution to the stellar metallicity problem is through early enrichment of the gas that contributes to the formation of most of the 
BCG mass at high redshift. We stress that such form of enrichment could be missed by the R4K and R8K simulations because of resolution limitations: if the 
resolution of the simulations is too low to capture early star formation in small haloes, the effect of their metal enrichment will be missing from the rest 
of the simulations. {However, comparison between R4K and R8K shows that this effect is mild. At $z=0.5$ the metallicity slightly increases passing from R4K to R8K. From our results, it is yet unclear if and 
how much the stellar metallicity will increase as the resolution is further increased. } A more detailed discussion of the effects of resolution on setting the stellar metallicity can be found in Section~\ref{sec:resolution_effects}.

\subsection{Stellar properties in cool cores and non-cool core clusters}\label{sec:cc_ncc}

\begin{figure*}
\begin{center}
\includegraphics[width=0.99\textwidth]{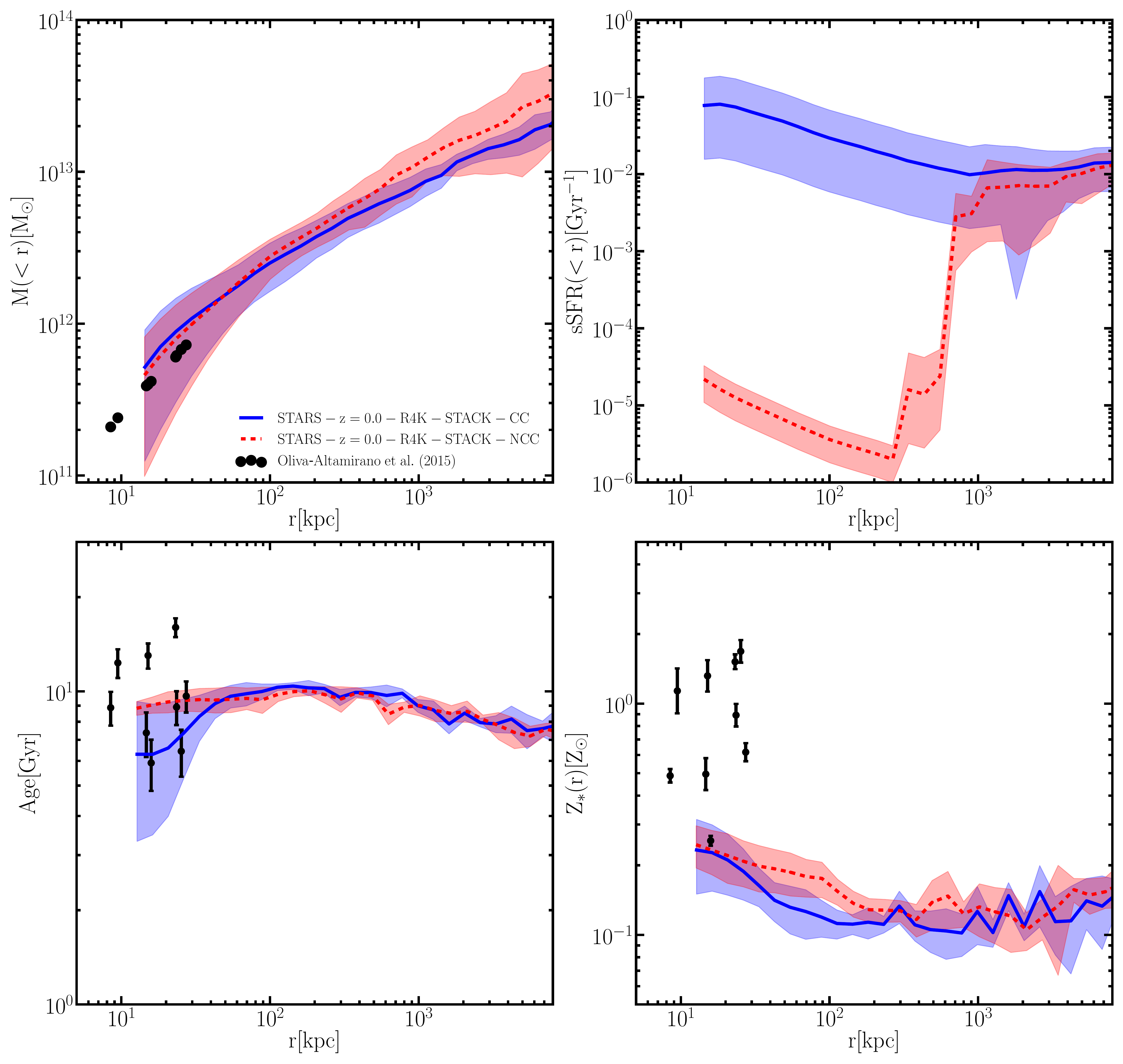}
\end{center}
\caption{\label{fig:stprop4k_cc_ncc}Stellar properties for the R4K runs at redshift $z=0$ for the cool core (solid blue line) and non-cool core (dashed red line) clusters. The minimum radius 
plotted for all the profiles is $r_{\rm min}=2\Delta x$, and the maximum radius is $r_{\rm max}\sim4-6R_{\rm 200,m}$ at high redshift and 8 Mpc at $z\leq 0.5$. Top left: cumulative stellar mass. 
Top right: specific star formation 
within a given radius averaged over a time interval $\Delta t=10^8$ yr. Bottom left: mean stellar age. 
Bottom right: stellar metallicity. In all panels the dark shaded areas 
represent the typical 1-$\sigma$ scatter among different haloes at redshift $z=0$. The observational data from the sample of BCGs at 
redshift $z<0.095$ of Oliva-Altamirano et al. (2015) is plotted (black points with error bars); their data is extrapolated to and plotted at twice the effective radius of the 
galaxy; the extrapolation is done using the gradients measured and reported in their paper. }
\end{figure*}

\cite{2015arXiv150904289H} discuss the cool core/non-cool core dichotomy in the {\sc Rhapsody-G} sample in detail, showing that the cool core nature cannot be removed by any 
of the thermal AGN feedback models we explored and that the state of the cluster depends on the merger history of the cluster. In our simulations cool cores can be converted 
into non-cool cores if the angular momentum of halo mergers is low; if the angular momentum is large, cool cores in the progenitor of a merger may never interact directly 
and may survive. The state of the cluster core influences how efficient the cluster centre is at forming stars. In Figure~\ref{fig:stprop4k_cc_ncc} we split our clusters 
into cool core and non-cool core systems as in \cite{2015arXiv150904289H} and analyse the stellar properties at redshift $z=0$. The top right panel of 
Figure~\ref{fig:stprop4k_cc_ncc} shows a huge difference in the cumulative sSFR profile of cool core and non-cool core clusters. Star formation is efficiently quenched in the 
non-cool core clusters; however the cool-core clusters have significant residual star formation at their centres. 

Given the cosmological origin of the cool cores in our simulations, 
it appears evident that it is harder to suppress cool cores only via the present implementation of AGN feedback. However, the results shown here and in 
\cite{2015arXiv150904289H} do not suggest that this is a big issue for recovering stellar masses for the centrals that match those of observed objects. In fact, 
the top left panel of Figure~\ref{fig:stprop4k_cc_ncc} shows a very weak dependence of the stellar mass profile on the cool core/non-cool core state of the cluster. 
The bottom left panel of Figure~\ref{fig:stprop4k_cc_ncc} shows that the stellar age is lower at the centre of cool core clusters, arguably an effect of the residual central 
star formation. Finally, the stellar metallicity profiles of non-cool core clusters are much flatter than in cool core clusters, as shown in the bottom right panel of Figure~\ref{fig:stprop4k_cc_ncc}. \cite{2015arXiv150904289H} showed that non-cool core clusters typically undergo dramatic merging events which also cause the redistribution of metals in a shallower profile. We will show that a similar trend is observed in the gas metallicity profiles (Section~\ref{sec:ICM_met_cc_ncc}).

\subsection{Caveats on the stellar metallicity distribution: central and satellite galaxies}\label{sec:resolution_effects}

The stellar metallicity profiles in Figures~\ref{fig:stprop4k} and \ref{fig:stprop8k} show only partial information about the stellar metallicity distribution 
in our simulated clusters. In particular, it shows only the average metallicity in spherical shells and does not explicitly show the stellar metallicity of galaxies. To place additional constraints on the results of our simulations we need to study the population of galaxies.

\begin{figure}
\begin{center}
\includegraphics[width=0.49\textwidth]{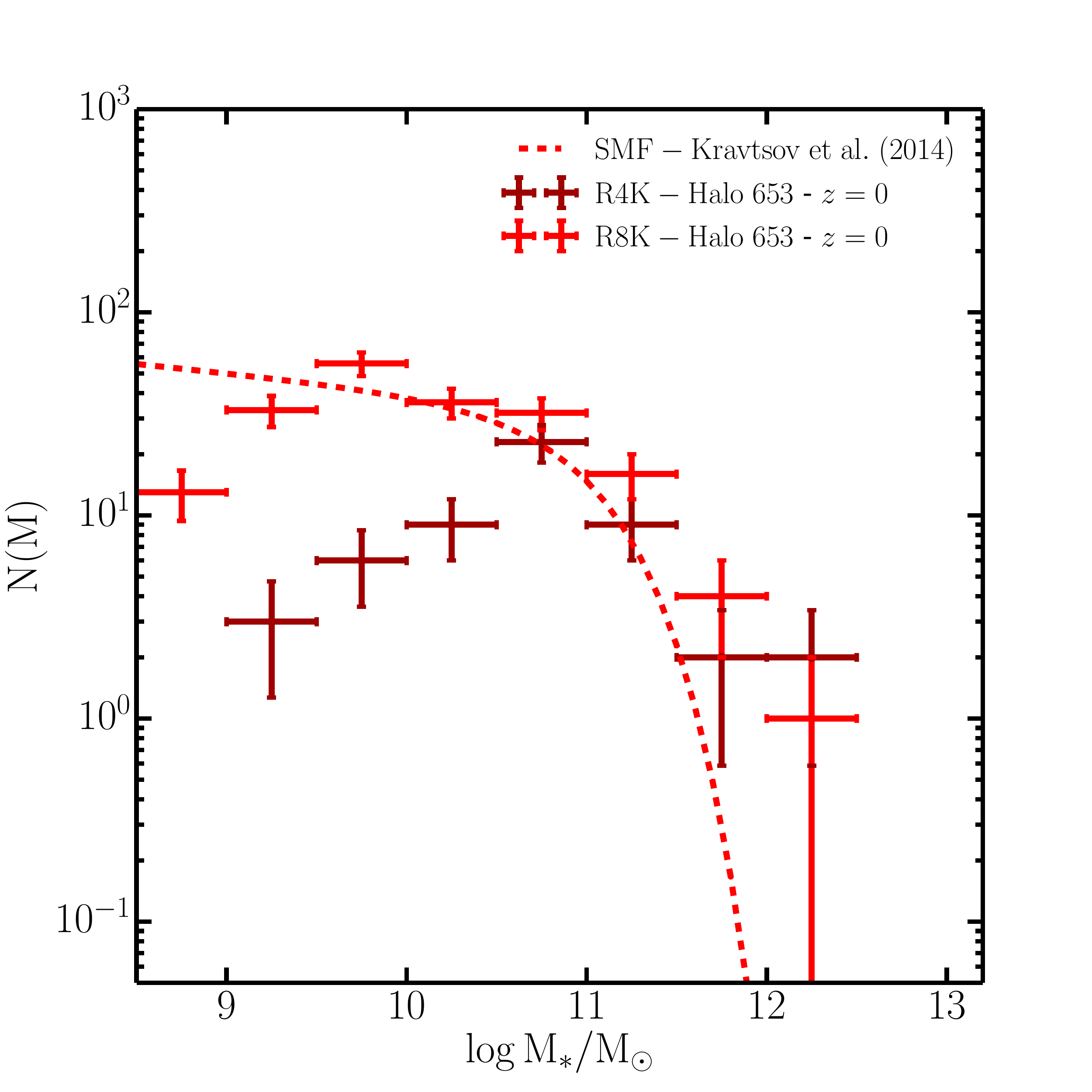}
\end{center}
\caption{\label{fig:smf} Satellite galaxy counts in bins of 0.5 dex in stellar mass for the satellites of halo 653. 
The red points with error bars represent the results of our simulations; the y-axis error bar is the Poisson scatter for the given counts, the x-axis error 
bar is the bin size used for the counts. The darker (brighter) points represent the R4K (8K) results. The red dashed line is the SMF in 
Kravtsov et al. (2014) and has been measured using galaxy counts in 8 clusters.}
\end{figure}

Only the most massive galaxies in the {\sc Rhapsody-G} sample are resolved with a large number of stellar particles. In the R4K run the stellar mass resolution 
only allows galaxies of stellar mass $\sim 5\times 10^{10}$ M$_{\odot}$ to be resolved with $\sim 1000$ particles. In the R8K runs this limit decreases by 
a factor 8, so a galaxy of mass $10^{10}$ M$_{\odot}$ can in principle be resolved with a few thousand particles. Unfortunately the statement is not true at 
all redshifts because of the multi-resolution nature of the simulations: the maximum mass resolution for the stellar particles is only activated at redshift 
$z<0.25$. This means that the older stellar populations will be resolved by a lower number of particles. One way to measure the completeness of our galaxy 
sample in a more empirical fashion is to compute the satellite galaxies stellar mass function (SMF) in our clusters and to compare it to observational 
measurements (Figure~\ref{fig:smf} for halo 653). As an observational constraint we use the Schechter form fit to the SMF of satellite galaxies in 8 clusters used by 
\cite{2014arXiv1401.7329K}. The simulations match the observationally determined SMF to a good degree of accuracy only within certain ranges of stellar mass: 
$10.5<\log M_{*}<11.5$ for R4K resolution and $9.5<\log M_{*}<11.5$ for R8K resolution. Below the minimum mass in these ranges the number of satellites starts declining. This is a hint that 
such populations of galaxies are not fully resolved. Galaxies of mass $\log M_{*}>11.5$ appear to be too massive by 0.5 dex in $\log M_{*}$ in the simulations. In our companion paper \citep{2015arXiv150904289H} we show that this is an effect of having slightly too high star formation activity in these objects. This is 
a manifestation of the fact that the current implementation of thermal AGN feedback does not provide enough heating to completely shut down star formation in 
such massive galaxies. 

Having a deficit (or excess) of galaxies compared to the measured SMF also means that there will be a deficit (or excess) of metals injected in the ICM and 
therefore in the newly formed stars. Even if the high end of the SMF is slightly overpredicted (due to slightly too high stellar masses), the least massive galaxies dominate in number and in stellar 
yield, so the simulations will have a metal deficit if the low mass population is not resolved. We try to quantify the amount of missing metals with a 
simple calculation; the discussion is somewhat technical and we defer the details to Appendix~\ref{appendix:A}, where we found that the metallicity from unresolved 
stellar populations can be up to $\sim 5\times 10^{-3}$ Z$_{\odot}$. We also point out that pre-enrichment models based on simulations of the formation of high redshift galaxies \citep[e.g.][]{2014ApJ...795..144C} can be used to take unresolved metal enrichment into account.

\begin{figure}
\begin{center}
\includegraphics[width=0.49\textwidth]{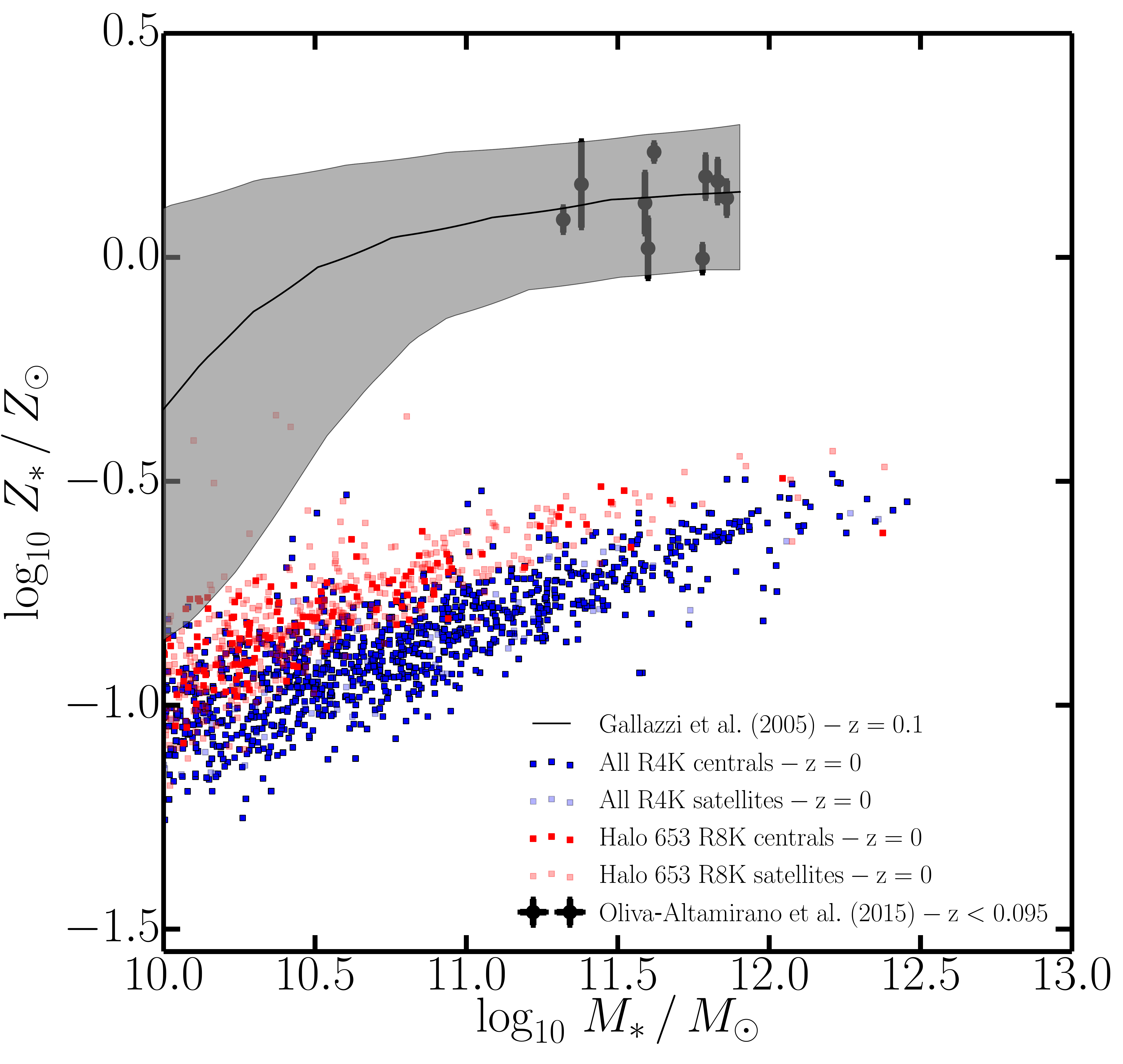}
\end{center}
\caption{\label{fig:mstar_zstar} Stellar mass-metallicity relation at low redshift for all the galaxies in the R4K runs (blue) and for all the galaxies 
in halo 653 R8K (red) runs. Centrals are represented by full colour squares, satellites by semi-transparent squares. The simulations are 
compared to the BCG data from Oliva-Altamirano et al. (2015; black points with error bars) and to the global stellar mass-metallicity relation determined 
by Gallazzi et al. (2005; black line with shaded area representing the 1-$\sigma$ scatter). }
\end{figure}

More constraints on the spatial distribution of metals can be achieved by looking at the stellar mass-metallicity relation of all the cluster galaxies, 
Figure~\ref{fig:mstar_zstar}. The stellar mass in this plot is the sum of the mass of all the stellar particles in the region in which the 3D stellar mass 
density is $\rho_{*}>2.5\times 10^6$ M$_{\odot}/$kpc$^3$ (as in \cite{Martizzi2012a}). The stellar metallicity in the plot is the mass-weighted mean metallicity in 
the same region and we treat it as a probe of the central metallicity of the galaxies. We find that the central metallicity of all galaxies in our clusters is 
$\sim 0.5$ dex lower than in the observations. The same result is found in both the R4K and R8K simulations. The results in this plot are in agreement with those in Appendix A of \cite{2014MNRAS.444.1453D} who analysed the Horizon-AGN simulation performed with {\sc Ramses} (at somewhat better resolution). 


As we have just discussed, the mass resolution of our simulations allows to resolve the mass function of galaxies in a well defined range of stellar mass, leaving the smallest galaxies sitting in the least massive dark matter haloes unresolved. However, 
the limited spatial resolution also places a constraint on the ability of resolving the internal structure of the cluster galaxies. As a matter of fact, if we take 
2 cell sizes as the effective softening of the gravitational force, its value is 9.5 kpc in R4K and 4.75 kpc in R8K. Such values for the softening are comparable 
to the effective radius of most galaxies \citep{2006ApJ...650...18T, 2007ApJ...670..206V, 2012MNRAS.419.3018C, 2014ApJ...788...28V}, so that their structure cannot 
be spatially resolved. As a result, the potential of most galaxies is too shallow and metals cannot be efficiently confined, so that most of them are ejected and mixed into the ICM. Increasing the spatial resolution will probably alleviate the problem. 

In Appendix~\ref{appendix:b}, we show that the value chosen for the metal yield $y$ can strongly influence the mass-metallicity relation. It should be stressed that the metal yield used in our simulations is the mean yield associated to a given stellar particle which represents an unresolved stellar population. Variation in the yield may depend on the environmental dependence of the initial mass function and may increase by a factor $\sim 4$ in massive clusters compared to the field \citep{2014MNRAS.444.3581R}. Furthermore, stellar winds and Type Ia SNe that return a large fraction of their stellar mass content back to the gas are not included in our simulations and would probably increase the effective yield (see e.g. \cite{2015MNRAS.446..521S}, \cite{2014MNRAS.444.1518V} for recent implementations in hydrodynamical codes). 

\cite{2011MNRAS.417.1853D} performed AMR simulations of clusters of smaller mass, obtained very similar results for their metallicity and reached very similar conclusions on the effects of resolution and higher stellar yields. The results of \cite{2011MNRAS.417.1853D} and our paper suggest that {\itshape by  combining higher $y$ and better resolution it should be possible to bring stellar metallicities up by a factor $\sim 4$ which might provide a much better match between simulations and observations.} 

However, it is important to stress that there is another discrepancy that is not likely to disappear simply by increasing the resolution and increasing the stellar yield. In fact, our simulations show a mass-metallicity relation with a unique slope, but the observed relation has a slope that depends on the stellar mass \citep{2005MNRAS.362...41G}, with the characteristic change in slope for stellar masses around $10^{10.7}M_{\odot}$. \cite{2013ApJ...772..119L} provide an excellent theoretical guideline to interpret the origin of the mass-metallicity relation. For any given stellar mass, the ratio between specific star formation rates and specific gas accretion rates onto the galaxies 
determines the mass-metallicity relation. This ratio changes as a function of stellar mass, producing the change of slope in the mass-metallicity relation. The stellar mass threshold beyond which most galaxies are quenched can be identified in the data of \cite{2005MNRAS.362...41G} as the stellar mass beyond which the mass-metallicity relation has a very shallow slope. {\itshape If the decline of specific star formation rate at masses larger than the quenching mass scale, the kink in the mass-metallicity relation will not be recovered properly.} We stress that such effect can be taken into account in analytical models by assuming a dependence of the specific star formation rate on the stellar mass as in \cite{2013ApJ...772..119L}. Our conclusion is that the change of slope in the mass-metallicity relation in our simulations is not recovered because of the residual star formation happening in some of the massive simulated galaxies, i.e. we do not properly recover the trend of specific star formation rates with stellar mass in massive galaxies and the observe quenched fraction at a given stellar mass (see Paper I). However, we note that this is notoriously a very hard  to reproduce these properties at a quantitative level: even state-of-the-art high resolution ($\lesssim0.5$ kpc) hydrodynamical cosmological simulations aimed at reproducing the properties of a large population of galaxies fail to recover the observed quenched fraction for a given stellar mass, with a tendency to over-produce star formation and stellar masses in the most massive sub-halos \citep{2014MNRAS.444.1518V, 2015MNRAS.452.2879T}. It is likely that more efficient schemes for feedback in massive galaxies will help solving this issue.

Despite the discrepancies found for the mass-metallicity relation, the satellite mass function produced by our simulations is relatively similar to the one determined from observations in the ranges $10.5<\log M_{*}<11.5$ for R4K, $9.5<\log M_{*}<11.5$ for R8K, and it is over-predicting the number of galaxies only when  $\log M_{*}>11.5$. This means that the total amount of metals produced by star formation in the simulations is probably not too far from the values expected from the observed satellite mass function. Thus, it is still very meaningful to analyse the metallicity of the ICM and to quantify the total amount of metals in the clusters. 

\section{Evolution of the gaseous content}

In this section we focus on the gaseous content of the {\sc Rhapsody-G} clusters. The accretion of gas and the evolution of the ICM metallicity is analysed in detail and differences between cool core and non-cool core clusters are highlighted. 

\subsection{ICM metallicity}

\begin{figure*} \begin{center} \includegraphics[width=0.49\textwidth]{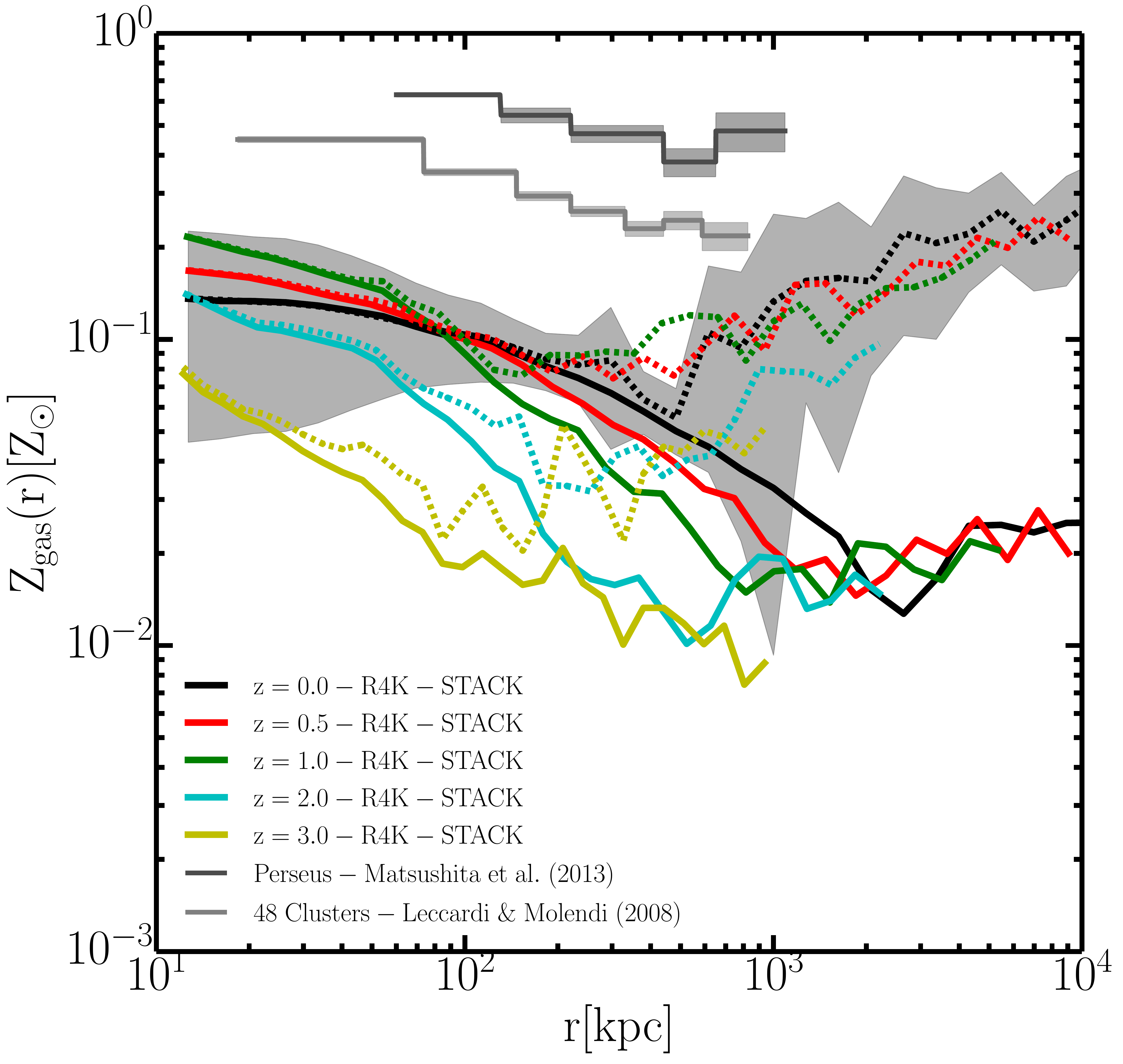} \includegraphics[width=0.49\textwidth]{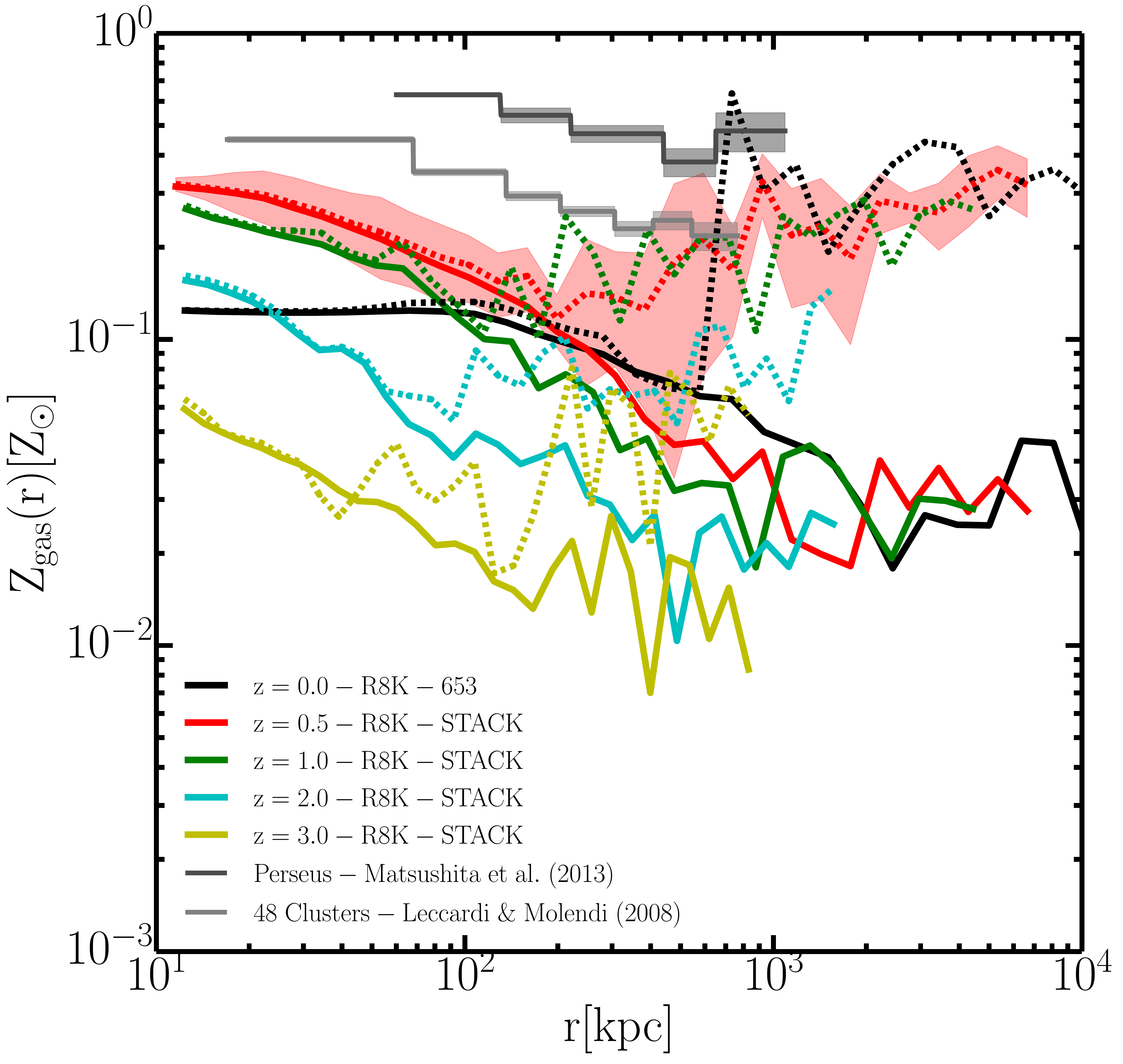} \end{center} \caption{\label{fig:metalstack} Differential metallicity profile for the R4K(left panel) and R8K (right panel) runs vs. observations. Solid coloured lines represent mass-weighted ICM metallicity at different redshifts, whereas dashed coloured lines represent X-ray emissivity weighted ICM metallicities.  The typical 1-$\sigma$ scatter among the clusters is represented by dark shaded area for R4K at redshift $z=0$ (left panel) and for R8K at redshift $z=0.5$ (right panel).  The simulations are compared to the observational results of Leccardi \& Molendi (2008; 48 clusters at redshift $0.1 \lesssim z\lesssim 0.3$ observed with XMM-Newton) and Matsushita et al. (2013; Perseus cluster observed with XMM-Newton) represented by grey lines with shaded areas representing the measurement errors.}
\end{figure*}

In this section, we discuss the evolution of the mass and metallicity profile of the {\sc Rhapsody-G} clusters. The focus will be on the effects we believe 
determine the metallicity content of the simulated clusters. First of all, we compare the metallicity profile of the simulated clusters to that of observed 
systems: in Figure~\ref{fig:metalstack} the differential ICM metallicity profiles of the {\sc Rhapsody-G} clusters are compared to data from the Perseus cluster 
observed with XMM-Newton \citep{2013ApJ...764..147M} and to data from 48 clusters at low redshift ($0.1 \lesssim z\lesssim 0.3$) selected by 
\cite{2008A&A...487..461L} from the XMM-Newton archive. {We choose these observational datasets because (I) the Perseus cluster is one of the most studied cool core clusters in the same mass range of our simulated sample and (II) the \cite{2008A&A...487..461L} sample contains a mixture of cool core and non-cool core clusters in a similar mass range, so that the average metallicity profile shown in Figure~\ref{fig:metalstack} can be thought as representative of the cluster population at low redshift. Even if the data from \cite{2008A&A...487..461L} is at redshift $z\sim 0.2$, we do not expect extreme variations of ICM metallicities with respect to redshift $z=0$ samples;  therefore, the comparison between this observational sample and our simulations is relevant, even with the small difference in redshift between data and simulations.} For the simulated clusters we differentiate between ICM metallicities computed by weighting each 
AMR cell by its gaseous mass (coloured solid lines) and ICM metallicities computed by weighting each cell by its X-ray emissivity (coloured dashed lines); for our purpose the X-ray emissivity is a weight $\propto \rho^2T^{1/2}$. The emissivity weighted metallicities agree 
very well with the mass-weighted values in the inner regions of the clusters; however, they are systematically higher at large radii. This effect is related to 
the fact that metals are distributed in relatively clumpy structures in the outer regions of the clusters. The effect of clumpiness can boost the emissivity 
weighted metallicity by almost a factor 10 in the outskirts of the clusters at redshift $z<1$. This fact by itself, might also imply that observations could 
provide strongly biased estimates of the metallicity. It is worth stressing that the emissivity weighted metallicities we measure in our simulations are all 
lower than the metallicities of the observed clusters. However, as we already observed in the previous section, the effect of resolution on metal yield from 
stars and therefore the stellar and ICM metallicity can be quite important. We see that the ICM metallicity at $z=0.5$ increases by a factor $\sim 1.5-2$ from 
the low resolution R4K runs to the higher resolution R8K runs, as observed for the stellar metallicity. The R8K runs seem to converge from below to the observed 
results. This fact suggests that with higher resolution (and better resolved star formation and metal yield from satellite galaxies) the correct metallicity 
of the clusters could be reproduced.

\subsection{Growth of the ICM mass profile and its effect on ICM metallicity}

\begin{figure*}
\begin{center}
\includegraphics[width=0.99\textwidth]{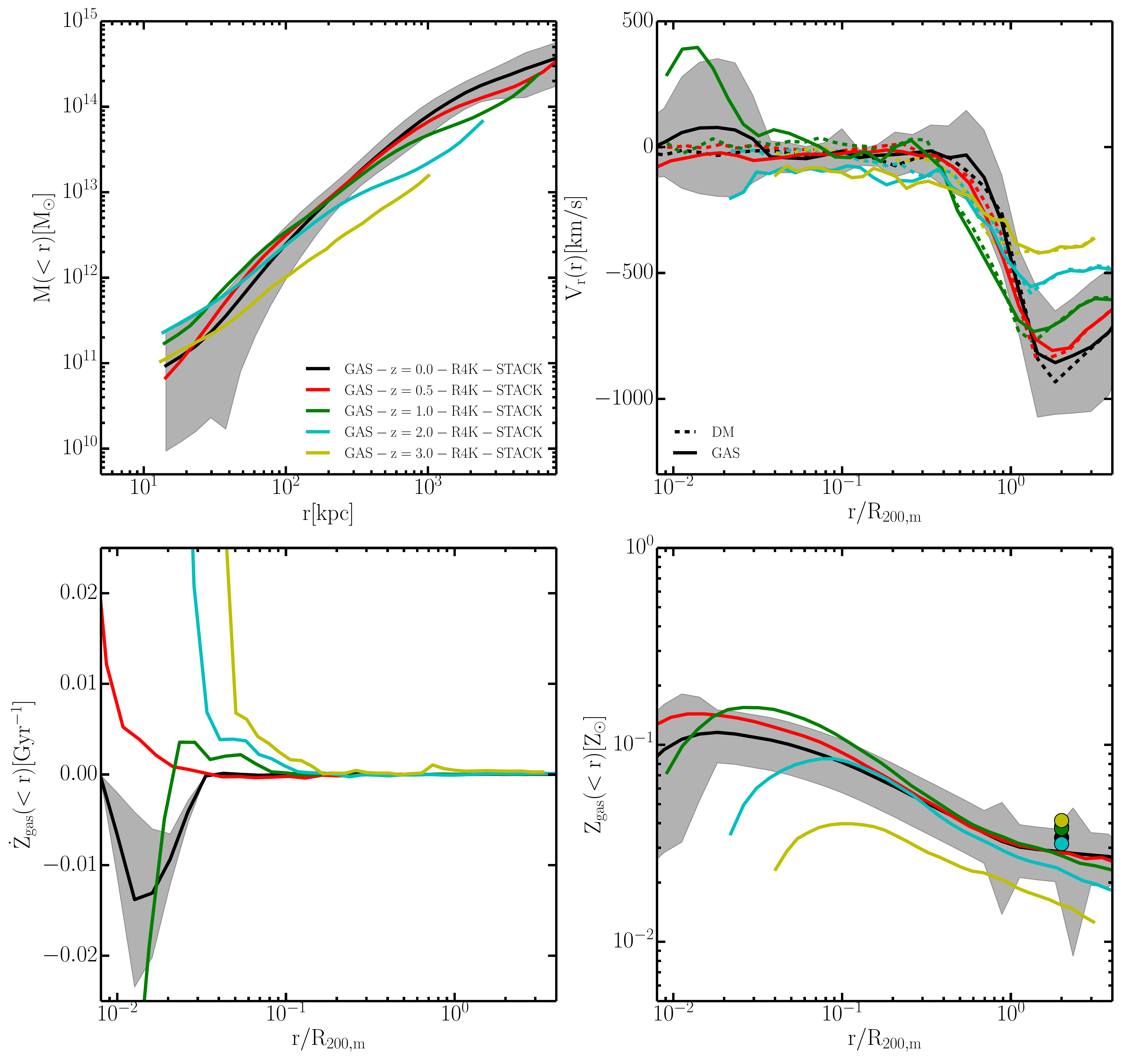}
\end{center}
\caption{\label{fig:accrate4k} Top left: gaseous mass profile for several redshifts (solid lines). Top right: radial velocity of dark matter (dashed lines) and gas mass (solid lines) at 
several redshifts; since the dark matter centre has been used the radial velocity of dark matter goes to zero at small radius, whereas the gas radial velocity does not; this choice does not
influence our results at large radii where our analysis is more relevant. Bottom left: derivative of the gas metallicity enclosed within a given radius $\dot{\rm Z}_{\rm gas}(<r)$. Bottom right: gas metallicity enclosed within a given radius $Z_{\rm gas}(<R)$; the coloured 
circles represent the predictions of the analytical model discussed in Section~\ref{sec:ana_mod}, with colours matching those of the solid lines for a given redshift. The dark shaded areas in all plots represent 
the typical 1-$\sigma$ scatter among different haloes at redshift $z=0$.}
\end{figure*}

\begin{figure*}
\begin{center}
\includegraphics[width=0.99\textwidth]{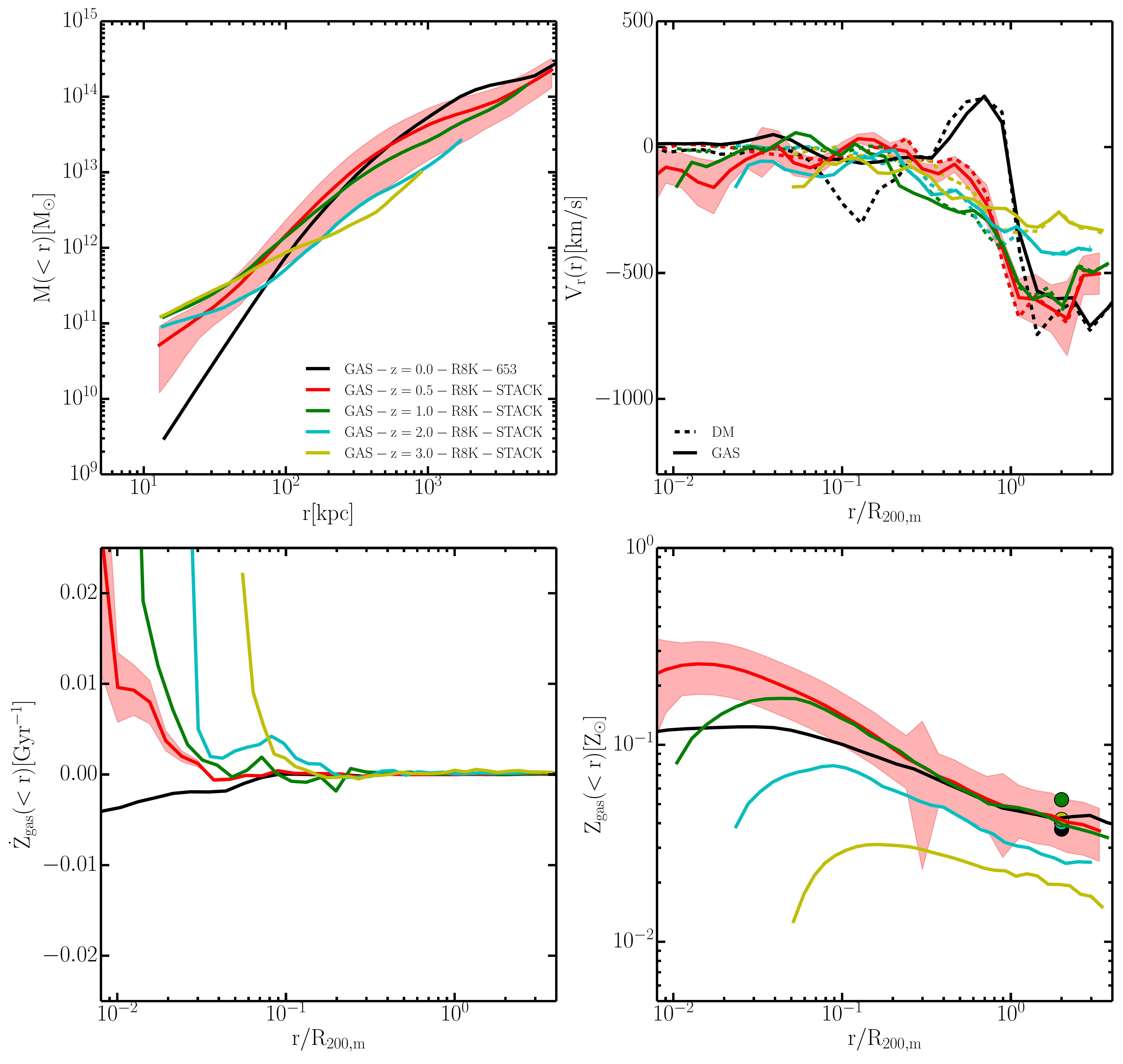}
\end{center}
\caption{\label{fig:accrate8k} Same as Figure~\ref{fig:accrate4k} but for the R8K. The redshift $z=0$ line is available only for halo 653 (black lines). The red shaded areas in all plots represent 
the typical 1-$\sigma$ scatter among different haloes at redshift $z=0.5$.}
\end{figure*}

In the previous section we discussed how the total amount of metals is a more robust quantity to study in these simulations, since it will more quickly converge as the 
mass resolution is increased. In the rest of this section we will focus on discussing the elements that determine the integrated value of the ICM metallicity of the cluster. 
First of all, we examine the gaseous mass profile of the clusters in the top left panels of Figure~\ref{fig:accrate4k} (R4K) and Figure~\ref{fig:accrate8k} (R8K). The figures 
show how the assembly of the gas distribution proceeds as a function of redshift. The very central regions ($r\lesssim 50$ kpc) of the clusters are the most affected by variable 
events like mergers and AGN activity. The gaseous mass distribution is more centrally concentrated at redshift $z>0.5$ and becomes less concentrated at redshift $z<0.5$. 
The gaseous distribution in the external regions of the clusters ($r\gtrsim100$ kpc) grows inside-out. This fact is explicitly demonstrated by the right panels of Figure~\ref{fig:accrate4k} (R4K) and Figure~\ref{fig:accrate8k} which show the spherically averaged 
radial velocity of dark matter (dashed lines) and gas (solid lines). The radial velocity plotted in this figure represents the net flow: 
positive values represent net outflowing motion, whereas negative values represent inflow. The x-axis shows 
the radius $r$ divided by $R_{\rm 200,m}$ at each redshift. The central regions of the clusters are usually characterised by decoupled motions of dark matter and gas. This region is 
very close to the central galaxy and the properties of the gas flow are influenced by the dynamics of infalling satellites and by the effect of feedback and cooling on the 
gas accretion mode \citep{2005MNRAS.363....2K, 2006MNRAS.368....2D, 2008MNRAS.390.1326O, 2009Natur.457..451D, 2011MNRAS.414.2458V, 2015MNRAS.448...59N}. On the other hand at the 
outskirts of the clusters the dark matter radial velocity is well coupled to the gas accretion velocity at all redshifts. The peak infall velocity increases with redshift as the 
halo grows in size and mass. Our results seem to confirm that the peak of the accretion velocity is always achieved at $r\sim 2R_{\rm 200,m}$ independently of redshift; this radial 
scale has been argued to be the typical scale at which the flow of matter onto haloes decouples from the Hubble flow 
\citep{2014arXiv1412.0662W, 2015arXiv150405591M, 2015ApJ...806...68L}. The presence of a gas inflow at $R_{\rm 200,m}<r< 2R_{\rm 200,m}$ at $z<3$ is an extremely relevant element for 
determination of the metal content of the cluster: if gas with lower metallicity is constantly accreted, and if metal mixing is efficient, this would tend to lower the average metallicity 
of the cluster. Of course this phenomenon will be contrasted by the metal yield coming from star forming activity within the cluster volume. The balance between these two effects 
determines the average metallicity of the cluster.

To quantify the variations of metallicity as a function of time, we measure the enclosed metallicity profile ${\rm Z}_{\rm gas}(<r)$ and its derivative $\dot{\rm Z}_{\rm gas}(<r)$; the enclosed metallicity profile yields the average ISM metallicity within a given radius $r$. $\dot{\rm Z}_{\rm gas}(<r)$ is plotted in the lower left panels of Figure~\ref{fig:accrate4k} (R4K) and Figure~\ref{fig:accrate8k} (R8K). It emerges from this plot that the time derivative of the metallicity can be quite high in the central regions of the halo at all redshifts. The higher the redshift, the higher the fraction of the volume within $R_{\rm 200,m}$ that is affected by variability. However, at large radius the metallicity derivative deviates from $\dot{\rm Z}_{\rm gas} \sim 0$ only at $z\gtrsim 2$; at all lower redshifts, the metallicity enclosed within $2R_{\rm 200,m}$ is approximately a constant. This means that metal yield from stars is effectively balancing the effect that accretion of (relatively) pristine gas has on the average metallicity.

The lower right panels of Figure~\ref{fig:accrate4k} (R4K) and Figure~\ref{fig:accrate8k} (R8K) explicitly show ${\rm Z}_{\rm gas}(<r)$. Being an integrated quantity, 
the enclosed metallicity profile differentiates from the differential profile of Figure~\ref{fig:metalstack} in the fact that it is smoother and that emissivity 
weighting is not applied; i.e. the metallicity at large radii is not boosted by clumping effects. This plot explicitly confirms that the average metallicity measured 
within large radii evolves very weakly with time. The coloured circles at $r=2R_{\rm 200,m}$ represent the prediction of a simple analytical model 
(Section~\ref{sec:ana_mod}) which assumes steady state for the enclosed metallicity at $r=2R_{\rm 200,m}$, i.e. $\dot{\rm Z}_{\rm gas}(<2R_{\rm 200,m})=0$; it is 
evident that the assumption of steady state reproduces quite well the results of the simulations. 

The main results of this section are that: (I) the differential ICM metallicity profile is closer to observations when emissivity weighting is used; (II) the simulations 
seem to converge to the results from the observations from below; however higher yields and still higher resolution than adopted here 
is required for proper convergence; (III) the mean 
metallicity of the simulated clusters is determined by the ratio of metal yield from star formation and accretion of metals from the IGM.

\subsection{ICM metallicity in cool core and non-cool core clusters}\label{sec:ICM_met_cc_ncc}
\begin{figure}
\begin{center}
\includegraphics[width=0.49\textwidth]{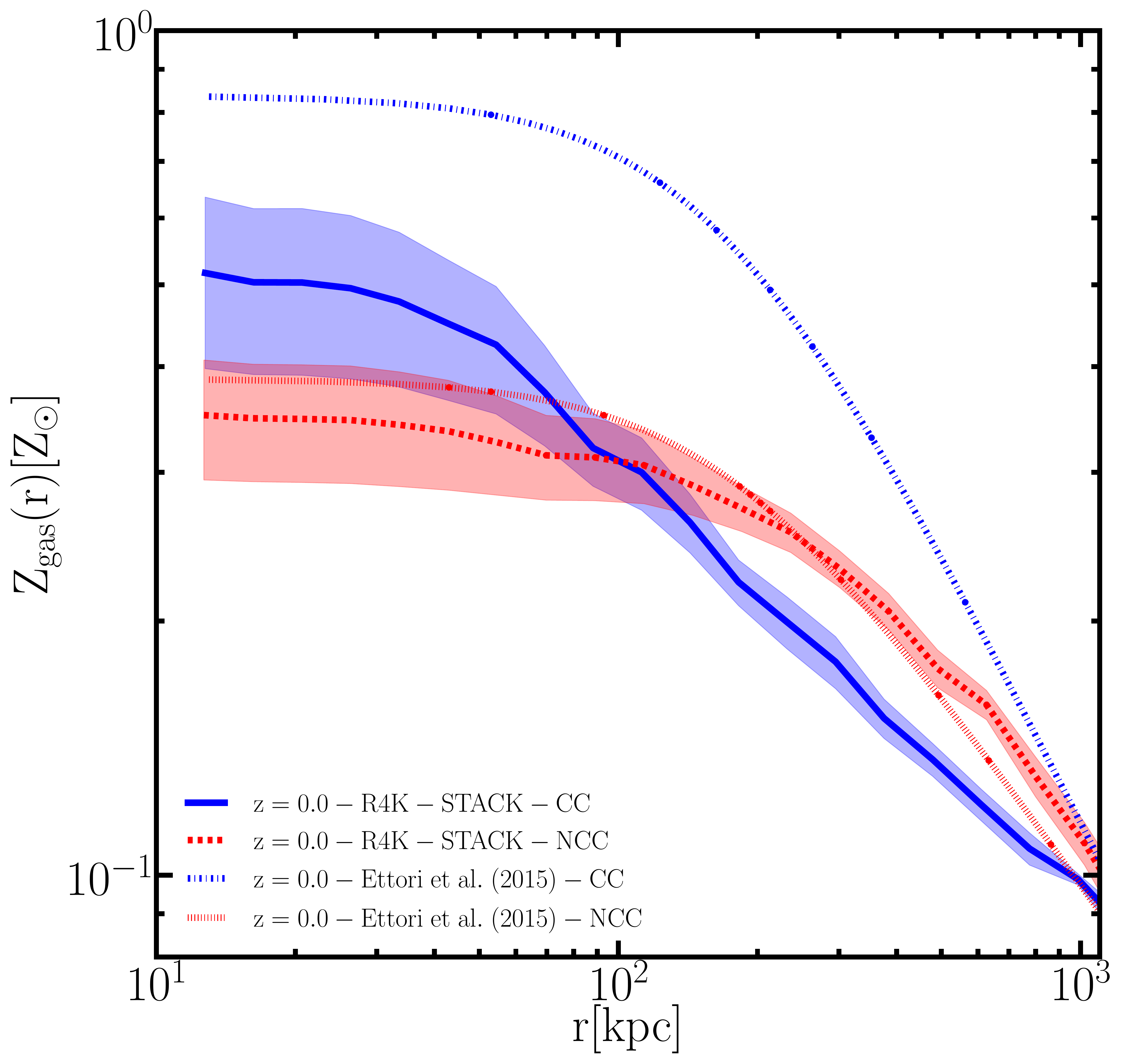}
\end{center}
\caption{\label{fig:metalstack_cc_ncc} Differential metallicity profile for the R4K cool core (blue solid line) and non-cool core (red dashed line) simulated clusters vs. observations. Solid coloured lines 
represent mass weighted ICM metallicity at different redshifts. The metallicities have been multiplied by a factor 3 to facilitate comparison to observations. The typical 1-$\sigma$ scatter among 
the haloes is represented by the blue and red shaded areas for the cool core and non-cool core clusters, respectively. 
The simulations are compared to the fits to observational data for CC (blue dot-dashed line) and NCC (red dotted line) clusters from Ettori et al. (2015). }
\end{figure}

In a recent paper, \cite{2015arXiv150904247R} showed that simulated cool core and non-cool core clusters have different ICM metallicity profiles, in agreement with observations. Figure~\ref{fig:metalstack_cc_ncc} shows the differential ICM metallicity profile for the R4K simulations at redshift $z=0$ when the cluster population is divided in cool core and non-cool core 
clusters as in \cite{2015arXiv150904289H}. {In this figure, the ICM metallicites have been multiplied by a factor 3 to facilitate comparison to the observational results of \citealt{2015A&A...578A..46E} (dashed lines; the sample is an extension of the one used by Leccardi \& Molendi 2008)}. Even if the scatter between the haloes (shaded areas) is large it is possible to appreciate how metals are re-distributed 
differently depending on whether the clusters have a cool core or not. As shown by \cite{2015arXiv150904289H}, cool cores can be converted into non-cool cores if the angular 
momentum of halo mergers is low; if the angular momentum is large, cool cores in the progenitor of a merger may never interact directly and may survive. 
Figure~\ref{fig:metalstack_cc_ncc} also shows that non-cool core clusters have lower metallicity and a flatter metallicity profile in the central regions. 
This fact is a consequence of the mixing that leads to flattened entropy profiles. {Comparison to the observational data of \cite{2015A&A...578A..46E} shows that our simulated clusters have metallicities a factor $\gtrsim 3$ smaller than real cluster both for CCs and NCCs; observations also show that the dichotomy between CCs and NCCs central metallicities is stronger in real clusters than in our simulated sample. In the SPH simulations by \cite{2015arXiv150904247R} this discrepancy 
is not observed; its origin in our simulations is unclear, however we are planning to investigate on this issue by performing explicit comparison of AMR and SPH simulations with the same initial conditions and similar sub-grid models in a future paper. }

\section{A simple model for the ICM metallicity evolution} \label{sec:ana_mod}

In this section, we develop a simple analytical model to predict the equilibrium metallicity of clusters and that is compared to the results of the simulations in the bottom right panels of 
Figure~\ref{fig:accrate4k} and Figure~\ref{fig:accrate8k}. The model describes a highly simplified scenario, it is only aimed at reproducing the value of 
the mean metallicity of the ICM within $r\sim 2R_{\rm 200,m}$ and is largely inspired by the `regulator' model discussed by \cite{2013ApJ...772..119L} and \cite{2015MNRAS.449.3274F}. 

Let us assume that the whole region $r<2R_{\rm 200,m}$ can be described with a 1-zone model. This statement is equivalent to saying that we will work with quantities integrated within this region, e.g. average star formation rate in the cluster, average metallicity, etc. Our goal is to model the effects that change the metallicity of the cluster. First of all, the metallicity of the ICM is defined as \begin{equation}
 {\rm Z}_{\rm gas}=\frac{M_{\rm Z}}{M_{\rm gas}}
\end{equation}
where $M_{\rm Z}$ and $M_{\rm gas}$ are the total metal mass and the total gaseous mass within the considered region, respectively. 

We want to evaluate the metallicity of the ICM when the system is in a steady state characterised by $d{\rm Z_{\rm gas}}/dt=0.$ The time derivative of the 
metallicity is then given by:
\begin{equation}\label{eq:metder}
 \frac{d{\rm Z_{\rm gas}}}{dt}=\frac{1}{M_{\rm gas}}\left[ \frac{dM_{\rm Z}}{dt}-{\rm Z_{\rm gas}}\frac{dM_{\rm gas}}{dt}\right].
\end{equation}
To compute the time derivative of the metallicity we need to compute the terms on the r.h.s. of eq.~(\ref{eq:metder}). 

The time derivative of the mass of metals in this region is given by 
\begin{equation}\label{eq:mmass}
 \frac{dM_{\rm Z}}{dt}=\left[\frac{dM_{\rm Z}}{dt}\right]_{\rm acc}+\left[\frac{dM_{\rm Z}}{dt}\right]_{\rm SF}
\end{equation}
where the first term on the r.h.s. is the rate of change generated by net accretion of metals from the IGM and the second term on the r.h.s. is the rate of 
change generated by star formation. 

Let us first consider the effect of star formation. Gas is converted into stars at a given star formation rate ${\rm SFR_{\rm true}}$. After each star formation event a fraction $\eta$ of the initial mass that collapsed into stars is ejected by winds and supernovae; we adopted $\eta=0.1$ in our simulations.  {We can also define the star formation rate of the mass effectively converted into stars (the fraction that is not ejected by SNe), ${\rm SFR}$}; this rate is easier to measure from simulations, so we will use this value in our treatment (contrary to what is done by \citealt{2013ApJ...772..119L}). From our definition it follows that 
\begin{equation}
 {\rm SFR_{\rm true}}=\frac{\rm SFR}{1-\eta},
\end{equation}
We assume a mass fraction $y$ of the ejecta is composed of metals, i.e. it represents the metal yield from star forming events. In our simulations we adopted $y=0.1$. 
The rate at which metals are ejected from star forming regions is then
$$
y\eta{\rm SFR_{\rm true}}=y\eta\frac{\rm SFR}{1-\eta}.
$$
Star formation does not only have the effect of adding metals. As gas is converted into stars, metals are consumed. The rate at which metals are consumed by star 
formation is
$$
-{\rm Z_{\rm gas}}(1-\eta)\times{\rm SFR_{\rm true}}=-{\rm Z_{\rm gas}}{\rm SFR}. 
$$
In the context of galaxy models there is an additional term that we should consider, i.e. the mass removal from outflows generated by stellar feedback. 
However, such outflows are not strong enough to eject material from the cluster $r<2R_{\rm 200,m}$ region, so they do not play any role in changing the average 
metallicity of the system. {Note that this assumption is valid only if a sufficiently large region is considered; in our case with $r\sim2R_{\rm 200,m}$ it is a safe assumption, but it will break down if smaller radii are considered.} 
The rate of change of the cluster metal mass due to star formation is given by the sum of the two terms we just discussed:
\begin{equation}\label{eq:mmass_sf}
 \left[\frac{dM_{\rm Z}}{dt}\right]_{\rm SF}=y\eta\frac{\rm SFR}{1-\eta}-{\rm Z_{\rm gas}}{\rm SFR}.
\end{equation}
{Note that the second term only takes into account the mass effectively converted into stars. }

The first term on the r.h.s. of eq.~(\ref{eq:mmass}) is related to the net flow of metals from the IGM:
\begin{equation}\label{eq:mmass_acc}
 \left[\frac{dM_{\rm Z}}{dt}\right]_{\rm acc}=-\oiint_{\Sigma_{\rm 200,m}}{\rm Z_{\rm IGM}}\phi_{\rm gas}d\Sigma,
\end{equation}
where $\Sigma_{\rm 200,m}$ is the surface of the sphere of radius $R_{\rm 200,m}$, ${\rm Z_{\rm IGM}}$ is the metallicity of the IGM and $\phi_{\rm gas}$ is the radial 
mass flux of gas per unit time; $\phi_{\rm gas}<0$ for net inflow and $\phi_{\rm gas}>0$ for net outflow. The integral can be directly evaluated from simulations.

Now what is left is to compute the derivative of ${M_{\rm gas}}$. This is very easily computed as the sum of the mass accretion rate from the IGM and the 
mass consumption rate due to star formation, i.e.:
\begin{equation}\label{eq:gmass}
 \frac{dM_{\rm gas}}{dt}=-\oiint_{\Sigma_{\rm 200,m}}\phi_{\rm gas}d\Sigma-{\rm SFR}.
\end{equation}
{Note that the second term only takes into account the mass effectively converted into stars, as it should be. In fact, the mass that is first incorporated into stars and then ejected by stellar feedback is quickly expelled and becomes available to the gas reservoir. }

\begin{figure*}
\begin{center}
\includegraphics[width=0.99\textwidth]{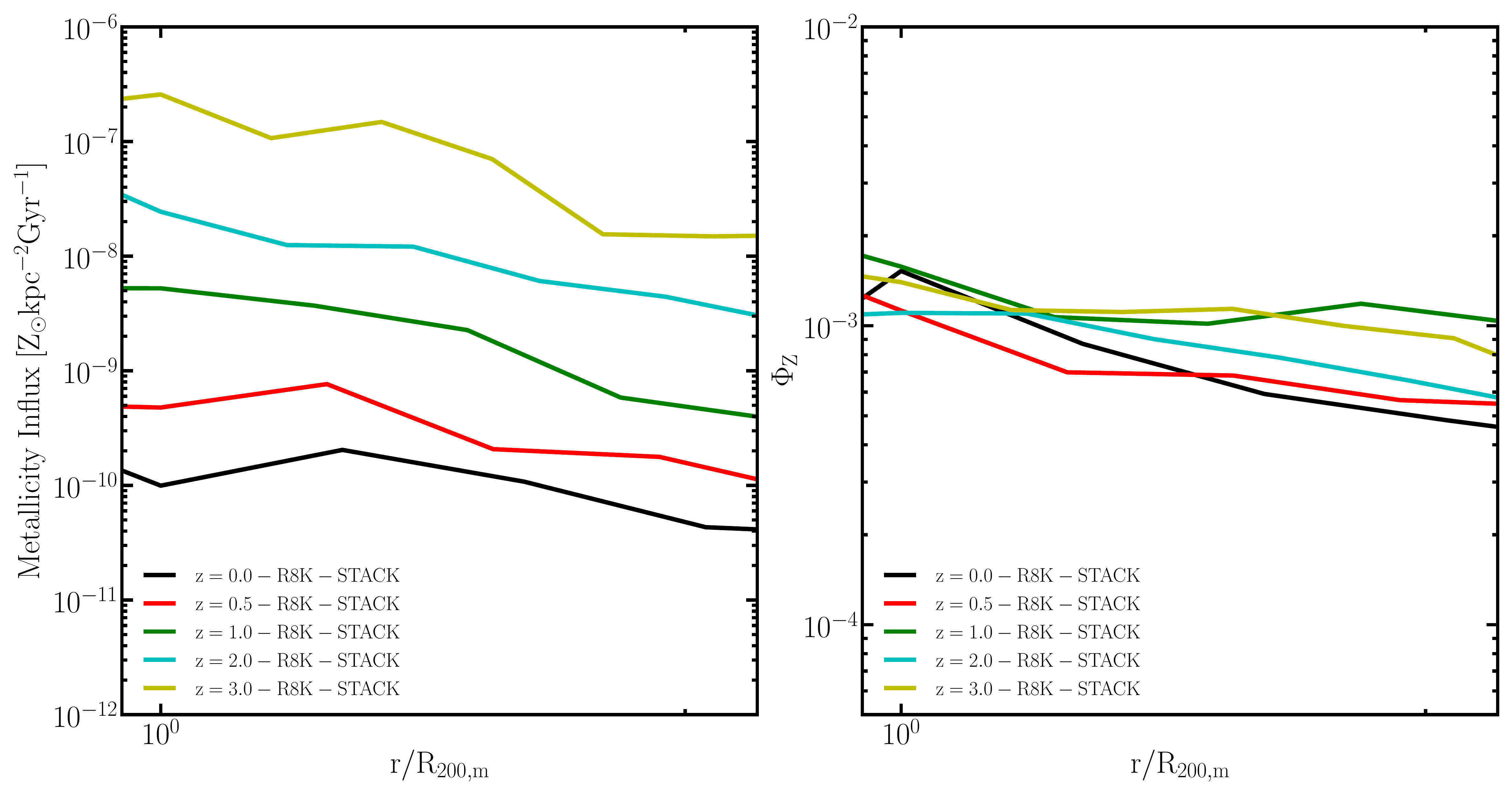}
\end{center}
\caption{\label{fig:anamod_fluxes} R8K simulations. Left: net metallicity influx across spherical surfaces centered at the cluster centre at different redshifts; positive values are for metal accretion. Right: the ${\rm \Phi}_{\rm Z}$ parameter which measures the balance between production of metals in the cluster and accretion of gas from outside the cluster at different redshifts. }
\end{figure*}

If we substitute eqs.~(\ref{eq:mmass}),( \ref{eq:mmass_sf}), (\ref{eq:mmass_acc}) and (\ref{eq:gmass}) into eq.~(\ref{eq:metder}) we obtain an explicit expression 
for the metallicity evolution 
\begin{equation}\label{eq:metder_final}
  \frac{d{\rm Z}_{\rm gas}}{dt}=\frac{1}{M_{\rm gas}}\left[ y\eta\frac{\rm SFR}{1-\eta} + \oiint_{\Sigma_{\rm 200,m}} ({\rm Z_{\rm gas} - Z_{\rm IGM}})\phi_{\rm gas} d\Sigma \right].
\end{equation}
If we set $d{\rm Z}_{\rm gas}/dt=0$ and solve for ${\rm Z_{\rm gas}}$, we get the steady-state value of the ICM metallicity:
\begin{equation}\label{eq:meteq}
 {\rm Z}_{\rm gas,eq}={\rm Z}_{\rm IGM}-\frac{y\eta {\rm SFR}}{(1-\eta) \oiint_{\Sigma_{\rm 200,m}}\phi_{\rm gas} d\Sigma },
\end{equation}
where we have assumed that ${\rm Z}_{\rm IGM}$ is approximately constant on the surface we are considering. All the quantities on the r.h.s. of eq.~(\ref{eq:meteq}) can be measured from the simulations and the equilibrium value can be compared to the actual metallicity 
to assess whether the assumption of steady state is valid or not. We label the second term on the r.h.s of eq.~\ref{eq:meteq} as
\begin{equation}
{\rm \Phi}_{\rm Z}= \left|\frac{y\eta {\rm SFR}}{(1-\eta) \oiint_{\Sigma_{\rm 200,m}}\phi_{\rm gas} d\Sigma }\right|.
\end{equation}
As we already mentioned in the previous section, the approximation of steady state is quite accurate at $z<3$ within a sphere of radius $2R_{\rm 200,m}$. Figure~\ref{fig:anamod_fluxes} demonstrates this fact explicitly: the left panel shows that the metallicity influx across spheres of radius $r\sim2R_{\rm 200,m}$ for the 8K simulations changes by several orders of magnitude across different redshift. However, this constant evolution of the metallicity influx is compensated by the increase of the total star formation rate in the cluster, so that the ratio ${\rm \Phi}_{\rm Z}$ measured from the R8K simulations always keeps the same order of magnitude (right panel of Figure~\ref{fig:anamod_fluxes}), i.e. the metallicity within a sphere of radius $2R_{\rm 200,m}$ is in a quasi-steady state which can be approximated by eq.~(\ref{eq:meteq}).

We note that the metallicity in eq.~(\ref{eq:meteq}) can be increased by either increasing the metal yield or reducing the net gaseous inflow onto the cluster. The latter could be achieved by either preventing cosmological gas accretion or by ejecting large quantities of gas from the clusters. However, producing such strong ejection events from clusters would require significant reduction of the baryon fraction to unrealistically low values. As we have demonstrated in paper~I, the current implementation of the feedback model (which is very similar to \citealt{2009MNRAS.398...53B}), is unable to appreciably affect gas at large scales, so that the mechanism to achieve such an effect is yet to be found.

\section{Comparison to recent simulations}

The distribution of metals in the most massive haloes in the universe and in satellite galaxies in cosmological hydrodynamical simulation that include prescriptions for AGN feedback has been studied in a series of recent works. It is important to compare our results to those of other authors to highlight the differences and to learn about the limits of current models of galaxy formation. 

A variety of semi-analytical models have been used to study metal enrichment in galaxy clusters with the conclusion that inclusion of ram-pressure stripping, galactic winds, metal dependent cooling and AGN feedback greatly improve the match between the metallicity profiles predicted by the models and those inferred from observations \citep{2008MNRAS.386...96C, 2009A&A...504..719K, 2010ApJ...716..918A, 2013MNRAS.428.1225S}. Cosmological hydrodynamical simulations have been used with similar goals \citep{2003MNRAS.339.1117V, 2006MNRAS.371..548R, 2007MNRAS.382.1050T, 2008MNRAS.391..110D, 2011MNRAS.415..353W, 2013MNRAS.432.3005C, 2013ApJ...763...38S}: the results by \cite{2007MNRAS.382.1050T} showed good agreement of metal abundances in simulated clusters in absence of AGN feedback at the price of significant over-production of galaxy stellar masses. More recent papers highlighted that much better agreement between stellar masses and cluster metallicities with observational results can be achieved when AGN feedback is included \citep{2011MNRAS.415..353W, 2013MNRAS.432.3005C, 2013ApJ...763...38S}. For this reason, in the following discussion we focus on a comparison to the most recent cosmological hydrodynamical simulations of clusters that include AGN feedback.

As already mentioned, the results we have shown so far are in good agreement with the results found by \cite{2014MNRAS.444.1453D} in the Horizon-AGN simulation and in previous zoom-in simulations performed with {\sc Ramses} \citep{2011MNRAS.417.1853D}.

\cite{2011MNRAS.415..353W} analysed the metal distribution in the ICM in the OWLS suite of simulations, which were run with the SPH code {\sc gadget3} \citep{2005MNRAS.364.1105S}. In their highest resolution versions, these simulations achieve a better mass ($\sim 6\times 10^6$ M$_{\odot}$) and spatial resolution (0.5 kpc/h) compared to the {\sc Rhapsody-G} runs in our paper. The authors analyse the average metallicity of the ICM in the cosmological boxes and in the range of redshift $0<z<3$ their results show qualitative agreement with the ICM metallicities we find in our simulations. However, their mean stellar metallicities appear to be higher than in our simulations and in better agreement with observations. There are two possible effects that might explain this discrepancy. First, the smaller value for the gravitational softening might allow galaxies to retain a larger fraction of the metals produced by supernovae; therefore increasing the local ISM metallicity and increasing the amount of metals in newly formed stars. Second, unlike AMR codes (and grid codes in general), standard SPH does not allow exchange of metals between resolution elements, leading to absence of metal mixing; again this goes in the direction of keeping a larger fraction of the metals bound to galaxies. Such differences, higher resolution and different metal mixing properties, might be extremely relevant for properly reproducing the stellar mass-metallicity relation; e.g. \cite{2015arXiv150708281S} discuss how the ejection and recycling of material is extremely relevant for setting the mass-metallicity relation. Indeed, the results from the EAGLE simulation (which uses an evolution of the OWLS sub-resolution models and was also ran with {\sc gadget3}) seem to confirm the results on the metal content of galaxies found in previous SPH simulations \citep{2015MNRAS.446..521S}. 

Interestingly, while the AGN feedback model adopted in OWLS/EAGLE is extremely similar to the one used in {\sc Rhapsody-G}, we do not find similar results when varying the parameters of the model \citep{2015arXiv150904289H}. In particular, when we vary the minimum temperature associated to AGN blasts $T_{\rm min}$ (equation~\ref{eq:min_ene}), we do not get a quasi-continuous change of the properties of the simulated clusters as a function of $T_{\rm min}$, which is observed in SPH simulations \citep{2014MNRAS.441.1270L}. We find that for $T_{\rm min} \leq 10^7$ K the metallicity of the ICM/stars does not vary significantly; if $T_{\rm min} > 5\times 10^7$ K metallicities are smaller by 20\%. Such differences in the behaviour of the same AGN feedback model in SPH and AMR suggest that sub-grid models couple differently in different schemes for hydrodynamics, a fact that may imply non-trivial consequences on the robustness of the results found by either scheme. 

Recently, \cite{2014MNRAS.438..195P} used {\sc gadget3} simulations with a different implementation of sub-resolution physics to study the chemo-dynamical properties of simulated galaxy clusters. These simulations seem to show results that differ significantly from what we found in {\sc Rhapsody-G}. In the galaxy group range $M_{\rm 500}<10^{14}$ M$_{\odot}$ the ICM metallicities are in good agreement with observational results; however they are a factor $\sim 3-4$ higher than in observations in the cluster range $M_{\rm 500}>10^{14}$. The mass and spatial resolution of these simulations is comparable to the one of the {\sc Rhapsody-G} runs and the sub-resolution modelling of AGN feedback is very similar to the one adapted for {\sc Rhapsody-G}: comparison of  the \cite{2014MNRAS.438..195P} results to {\sc Rhapsody-G} exemplifies the tension between AMR and SPH simulations of galaxy clusters: at fixed resolution, SPH and AMR seem to produce the opposite effect on ICM metallicities. 

Finally, it is worth mentioning the results achieved with the moving mesh code {\sc Arepo} \citep{2010MNRAS.401..791S} in the Illustris cosmological hydrodynamical simulation \cite{2014MNRAS.444.1518V}. The Illustris simulation evolved a cubic cosmological volume of side $\sim 100$ Mpc with dark matter mass resolution $\sim 6 \times 10^6$ M$_{\odot}$ and a gravitational softening of 0.7 kpc; the resolution is significantly better than in the {\sc Rhapsody-G} runs, but the volume is not large enough to have a significant number of clusters, so that a direct comparison to our results is not possible. Similarly to AMR codes, {\sc arepo} is also based on Riemann solvers and allows for explicit exchange of metals across resolution elements; this fact should bring the metal mixing properties of the code closer to the ones of AMR codes than to SPH codes. The Illustris simulations produce stellar metallicities in better agreement to the observations, probably as a result of the higher resolution and smaller gravitational softening that allows to better resolve the structure of galaxies and helps them to retain the metals produced by SNe. However, as shown by \cite{2014MNRAS.445..175G}, the galaxy clusters in Illustris have baryon fractions that are a factor $\sim 2$ lower than observed systems. This effect is the result of the particularly violent AGN feedback produced by the scheme implemented in this simulation that results in baryon depletion from the most massive haloes. In such a case, eq.~(\ref{eq:meteq}) predicts that this should result in a higher ICM metallicity in the cluster. 

The conclusion from this comparison is that differences in resolution, the hydrodynamical solvers, and in the implementations of sub-resolution models for AGN feedback all play a key role in determining the chemo-dynamical evolution of clusters and their galaxies. From the point of view of AMR simulations, we stress that simulations with significantly improved resolution and better AGN feedback prescriptions are needed to achieve better agreement with observations. 

\section{Summary and discussion}
We have performed a series of zoom-in simulations of very massive galaxy clusters ($M_{\rm vir}\approx 6\times 10^{14}$ M$_{\odot}/h$) performed with the AMR code {\sc Ramses} \citep{2002A&A...385..337T}. All the simulations account for gas cooling, the effect of an homogeneous photoionising UV background, star formation, supernova feedback and AGN feedback. \cite{2015arXiv150904289H} showed 
that these simulations are able to reproduce two distinct populations of cool core and non-cool core clusters and that the mass distribution, SZ properties and stellar content of the galaxies largely match currently available observational results. In this paper, we focus on the evolution of the stellar population, as well as the chemodynamical properties of both the stars and the intracluster medium over cosmic time.

Our analysis lead to several conclusions that we can summarise in the following points:

\begin{itemize}
\item The stellar mass in the central regions compares well to that measured for brightest cluster galaxies \citep{2015MNRAS.449.3347O}, but the modest ($\sim 5$ kpc) spatial resolution of the simulations artificially enhances tidal stripping, leading to a massive extended stellar halo.
 \item The specific star formation rate averaged within the entire cluster volume (density contrasts of a few hundred with respect to critical) is relatively low ($\sim 10^{-2}$ Gyr$^{-1}$, but the population mean increases toward the centre.  This increase is due to the subpopulation of cool core clusters in our sample; non-cool core clusters are quenched within the inner 300 kpc, or density contrasts of a few thousand.  In the outer regions dominated by infalling material, some of the satellite and central galaxies have residual star formation, i.e. are not totally quenched, driving the average upward, albeit weakly, with radius.  The star formation rates of central galaxies in cool core systems are higher than observed, suggesting that improvements to spatial resolution and mode of feedback may be required to suppress star formation in the most massive clusters. 
 \item The mean age of the stellar population at redshift $z=0$ agrees with the observations of \cite{2015MNRAS.449.3347O}; the results are consistent with a scenario in which most of the stellar mass consists of an old-age ($\sim 10$ Gyr at redshift $z=0$) stellar population. 
 \item The stellar metallicities throughout the cluster are lower than in the observations, but yields are uncertain and higher values are produced when the resolution is increased. The missing metal yield from unresolved galaxies is estimated using a simple analytical argument to increase $Z$ by only a small amount, $\sim 0.01$. 
 \item The metallicities of satellite galaxies are a factor $\gtrsim 5 $ too low compared to observations. These metallicities have to be treated differently than the average stellar metallicity because they are influenced by the fraction of metals from stellar yield that the galaxies are able to retain. The low values most likely reflect the inability of satellites to retain metals due to the large value gravitational softening that makes the gravitational potential very shallow.  Higher spatial resolution potentially combined with higher adopted metal yields from star formation motivated by observations (Appendix~\ref{appendix:b}) should alleviate this discrepancy. 
 \item The clusters accrete gas from the intergalactic medium and grow inside out. The accretion speed reaches a maximum at a radius $~2R_{200,m}$ as found by previous authors \citep{2014arXiv1412.0662W, 2015arXiv150405591M, 2015ApJ...806...68L}. Within this radius the metallicity of the ICM is well described by an equilibrium model in which accretion of low metallicity gas from the intergalactic medium is balanced by the production of metals from star formation events.
  \item In the outer regions of clusters, luminosity-weighted metallicities are systematically higher than mass-weighted values because metals tend to cluster in locally overdense structures at large cluster-centric radii. 
 \item Cool core clusters have steeper central ICM and stellar metallicity gradients than non-cool clusters. This result is in agreement with \cite{2015arXiv150904247R}, who use completely independent simulation methods.
 \item Comparison to the results of simulations with similar implementations of galaxy formation recipes and comparable resolution published in the literature provides interesting insight into the limitations of current models. The SPH simulation performed by \cite{2014MNRAS.438..195P} show ICM metallicities slightly higher than the observed ones and stellar metallicities closer to the observations than the ones found in {\sc Rhapsody-G}. 
 We speculate that this is an effect of the different properties of metal mixing in SPH codes compared to AMR codes; however this statement should be tested by direct comparison. The effect of resolution can be appreciated when comparing to the EAGLE \citep{2015MNRAS.446..521S} and Illustris \citep{2014MNRAS.444.1518V} simulations which have higher resolution (but a low number of high mass clusters): higher resolution may help satellite galaxies at some degree to retain more of the metals produced by star formation events.
\end{itemize}

While it is possible that the {\itshape conundrum} discussed by \cite{2014MNRAS.444.3581R} may be solved by re-analysis of available data on massive clusters, it is more likely that the solution to the problem may require modifying stellar yields, which may imply that the initial stellar mass function is sensitive to environment, at least in the extreme cases realised by massive cluster progenitors. 

Another fact that might alleviate the discrepancies between our simulations and metallicity measurements is the fact that we assume a cosmic baryon fraction 0.18, while the latest result from Planck suggest a cosmic baryon fraction 0.146 \citep{2015arXiv150201589P}. For a fixed amount of star-formation and hence metal production, that will lead to slightly underestimating the metallicity of the ICM. 

Independently of the quality of the observational data and on the value of the cosmological parameters, the fact that the results of several simulation techniques provide very different results concerning the metallicity distribution in clusters is an important issue: it tells us that the phenomena that play a role in setting the cluster metallicity are not implemented in an algorithm-independent way yet and are still affected by resolution-dependent modelling. The discussion in Appendix~\ref{appendix:b} shows that the exact metal content of simulated clusters also depends on the specific choice for the value of the stellar yield. Better modelling demands future simulations to resolve the satellite galaxies better than is currently achieved, since they play a very important role in determining the net metal injection from star formation events. The issue of convergence with increasing resolution has to be carefully taken into account. Convergence is not simple to assess in these multi-scale simulations. In fact, just increasing the resolution might produce better numerically converged properties for galaxies, gas mixing, ram-pressure stripping {\it for a given set of sub-grid models}. However, such models rapidly become inaccurate and unphysical if the resolution is increased by several orders of magnitude. For these reasons, achieving simulations that converge to a physical solution at high resolution is extremely challenging and efforts have to be spent in (I) improving the numerical efficiency of the codes to achieve better resolution, and (II) developing better physically motivated models which have weaker dependencies on resolution. For state-of-the-art and future simulations of galaxy clusters at resolution better than $\sim$kpc, different and more sophisticated models of feedback should be considered if one is interested in properly suppressing the formation of excess mass in the most massive satellites and in the centrals. Such models would include improved models for AGN feedback as well as sub-grid modelling of turbulent and cosmic ray pressure. It is one of the goals of the {\sc Rhapsody-G} project to implement such improvements to better match observations and simulations in the future.

\section*{Acknowledgments}
We thank Peter Behroozi for making the {\sc Rockstar-Galaxies} software available for us. We would also like to thank Robert Feldmann for the useful discussions on the `regulator model'. D.M. acknowledges support from the Swiss National Science Foundation (SNSF) through the SNSF Early.Postdoc and Advanced.Postdoc Mobility Fellowships. O.H. acknowledges support from the Swiss National Science Foundation (SNSF) through the Ambizione fellowship. HW acknowledges support by the U.S. Department of Energy under contract number DE-FG02-95ER40899. RHW received support from the U.S. Department of Energy contract to SLAC no. DE-AC02-76SF0051. This work was supported by a grant from the Swiss National Supercomputing Centre (CSCS) under project ID s416.

\bibliography{references} 

\begin{thebibliography}{}

\bibitem[\protect\citeauthoryear{{Anders} \& {Grevesse}}{{Anders} \&
  {Grevesse}}{1989}]{1989GeCoA..53..197A}
{Anders} E.,  {Grevesse} N.,  1989, \gca, 53, 197

\bibitem[\protect\citeauthoryear{{Arieli}, {Rephaeli} \& {Norman}}{{Arieli}
  et~al.}{2010}]{2010ApJ...716..918A}
{Arieli} Y.,  {Rephaeli} Y.,    {Norman} M.~L.,  2010, \apj, 716, 918

\bibitem[\protect\citeauthoryear{{Begelman}, {Volonteri} \& {Rees}}{{Begelman}
  et~al.}{2006}]{2006MNRAS.370..289B}
{Begelman} M.~C.,  {Volonteri} M.,    {Rees} M.~J.,  2006, \mnras, 370, 289

\bibitem[\protect\citeauthoryear{{Behroozi}, {Wechsler} \& {Wu}}{{Behroozi}
  et~al.}{2013}]{2013ApJ...762..109B}
{Behroozi} P.~S.,  {Wechsler} R.~H.,    {Wu} H.-Y.,  2013, \apj, 762, 109

\bibitem[\protect\citeauthoryear{{Bernardi}}{{Bernardi}}{2009}]{2009MNRAS.395.1491B}
{Bernardi} M.,  2009, \mnras, 395, 1491

\bibitem[\protect\citeauthoryear{{Bleuler}, {Teyssier}, {Carassou} \&
  {Martizzi}}{{Bleuler} et~al.}{2014}]{2014arXiv1412.0510B}
{Bleuler} A.,  {Teyssier} R.,  {Carassou} S.,    {Martizzi} D.,  2014,
  arXiv:1412.0510

\bibitem[\protect\citeauthoryear{{Booth} \& {Schaye}}{{Booth} \&
  {Schaye}}{2009}]{2009MNRAS.398...53B}
{Booth} C.~M.,  {Schaye} J.,  2009, \mnras, 398, 53

\bibitem[\protect\citeauthoryear{{Bromm} \& {Loeb}}{{Bromm} \&
  {Loeb}}{2003}]{2003ApJ...596...34B}
{Bromm} V.,  {Loeb} A.,  2003, \apj, 596, 34

\bibitem[\protect\citeauthoryear{{Brough}, {Collins}, {Burke}, {Mann} \&
  {Lynam}}{{Brough} et~al.}{2002}]{2002MNRAS.329L..53B}
{Brough} S.,  {Collins} C.~A.,  {Burke} D.~J.,  {Mann} R.~G.,    {Lynam} P.~D.,
   2002, \mnras, 329, L53

\bibitem[\protect\citeauthoryear{{Burke} \& {Collins}}{{Burke} \&
  {Collins}}{2013}]{2013MNRAS.434.2856B}
{Burke} C.,  {Collins} C.~A.,  2013, \mnras, 434, 2856

\bibitem[\protect\citeauthoryear{{Burke}, {Hilton} \& {Collins}}{{Burke}
  et~al.}{2015}]{2015MNRAS.449.2353B}
{Burke} C.,  {Hilton} M.,    {Collins} C.,  2015, \mnras, 449, 2353

\bibitem[\protect\citeauthoryear{{Chan}, {Kere{\v s}}, {O{\~n}orbe}, {Hopkins},
  {Muratov}, {Faucher-Gigu{\`e}re} \& {Quataert}}{{Chan}
  et~al.}{2015}]{2015arXiv150702282C}
{Chan} T.~K.,  {Kere{\v s}} D.,  {O{\~n}orbe} J.,  {Hopkins} P.~F.,  {Muratov}
  A.~L.,  {Faucher-Gigu{\`e}re} C.-A.,    {Quataert} E.,  2015,
  arXiv:1507.02282

\bibitem[\protect\citeauthoryear{{Chen}, {Wise}, {Norman}, {Xu} \&
  {O'Shea}}{{Chen} et~al.}{2014}]{2014ApJ...795..144C}
{Chen} P.,  {Wise} J.~H.,  {Norman} M.~L.,  {Xu} H.,    {O'Shea} B.~W.,  2014,
  \apj, 795, 144

\bibitem[\protect\citeauthoryear{{Conroy}, {Wechsler} \& {Kravtsov}}{{Conroy}
  et~al.}{2007}]{2007ApJ...668..826C}
{Conroy} C.,  {Wechsler} R.~H.,    {Kravtsov} A.~V.,  2007, \apj, 668, 826

\bibitem[\protect\citeauthoryear{{Cooper}, {Griffith}, {Newman}, {Coil},
  {Davis}, {Dutton}, {Faber}, {Guhathakurta}, {Koo}, {Lotz}, {Weiner},
  {Willmer} \& {Yan}}{{Cooper} et~al.}{2012}]{2012MNRAS.419.3018C}
{Cooper} M.~C.,  {Griffith} R.~L.,  {Newman} J.~A.,  {Coil} A.~L.,  {Davis} M.,
   {Dutton} A.~A.,  {Faber} S.~M.,  {Guhathakurta} P.,  {Koo} D.~C.,  {Lotz}
  J.~M.,  {Weiner} B.~J.,  {Willmer} C.~N.~A.,    {Yan} R.,  2012, \mnras, 419,
  3018

\bibitem[\protect\citeauthoryear{{Cora}, {Tornatore}, {Tozzi} \&
  {Dolag}}{{Cora} et~al.}{2008}]{2008MNRAS.386...96C}
{Cora} S.~A.,  {Tornatore} L.,  {Tozzi} P.,    {Dolag} K.,  2008, \mnras, 386,
  96

\bibitem[\protect\citeauthoryear{{Crain}, {McCarthy}, {Schaye}, {Theuns} \&
  {Frenk}}{{Crain} et~al.}{2013}]{2013MNRAS.432.3005C}
{Crain} R.~A.,  {McCarthy} I.~G.,  {Schaye} J.,  {Theuns} T.,    {Frenk} C.~S.,
   2013, \mnras, 432, 3005

\bibitem[\protect\citeauthoryear{{Croton}, {Springel}, {White}, {De Lucia},
  {Frenk}, {Gao}, {Jenkins}, {Kauffmann}, {Navarro} \& {Yoshida}}{{Croton}
  et~al.}{2006}]{2006MNRAS.365...11C}
{Croton} D.~J.,  {Springel} V.,  {White} S.~D.~M.,  {De Lucia} G.,  {Frenk}
  C.~S.,  {Gao} L.,  {Jenkins} A.,  {Kauffmann} G.,  {Navarro} J.~F.,
  {Yoshida} N.,  2006, \mnras, 365, 11

\bibitem[\protect\citeauthoryear{{Dav{\'e}}, {Oppenheimer} \&
  {Sivanandam}}{{Dav{\'e}} et~al.}{2008}]{2008MNRAS.391..110D}
{Dav{\'e}} R.,  {Oppenheimer} B.~D.,    {Sivanandam} S.,  2008, \mnras, 391,
  110

\bibitem[\protect\citeauthoryear{{Dekel} \& {Birnboim}}{{Dekel} \&
  {Birnboim}}{2006}]{2006MNRAS.368....2D}
{Dekel} A.,  {Birnboim} Y.,  2006, \mnras, 368, 2

\bibitem[\protect\citeauthoryear{{Dekel}, {Birnboim}, {Engel}, {Freundlich},
  {Goerdt}, {Mumcuoglu}, {Neistein}, {Pichon}, {Teyssier} \& {Zinger}}{{Dekel}
  et~al.}{2009}]{2009Natur.457..451D}
{Dekel} A.,  {Birnboim} Y.,  {Engel} G.,  {Freundlich} J.,  {Goerdt} T.,
  {Mumcuoglu} M.,  {Neistein} E.,  {Pichon} C.,  {Teyssier} R.,    {Zinger} E.,
   2009, \nat, 457, 451

\bibitem[\protect\citeauthoryear{{Dubois}, {Devriendt}, {Teyssier} \&
  {Slyz}}{{Dubois} et~al.}{2011}]{2011MNRAS.417.1853D}
{Dubois} Y.,  {Devriendt} J.,  {Teyssier} R.,    {Slyz} A.,  2011, \mnras, 417,
  1853

\bibitem[\protect\citeauthoryear{{Dubois}, {Pichon}, {Welker}, {Le Borgne},
  {Devriendt}, {Laigle}, {Codis} \& {Pogosyan}}{{Dubois}
  et~al.}{2014}]{2014MNRAS.444.1453D}
{Dubois} Y.,  {Pichon} C.,  {Welker} C.,  {Le Borgne} D.,  {Devriendt} J.,
  {Laigle} C.,  {Codis} S.,    {Pogosyan} D. e.~a.,  2014, \mnras, 444, 1453

\bibitem[\protect\citeauthoryear{{Dubois} \& {Teyssier}}{{Dubois} \&
  {Teyssier}}{2008}]{2008A&A...477...79D}
{Dubois} Y.,  {Teyssier} R.,  2008, \aap, 477, 79

\bibitem[\protect\citeauthoryear{{Ettori}, {Baldi}, {Balestra}, {Gastaldello},
  {Molendi} \& {Tozzi}}{{Ettori} et~al.}{2015}]{2015A&A...578A..46E}
{Ettori} S.,  {Baldi} A.,  {Balestra} I.,  {Gastaldello} F.,  {Molendi} S.,
  {Tozzi} P.,  2015, \aap, 578, A46

\bibitem[\protect\citeauthoryear{{Ettori}, {Rasia}, {Fabjan}, {Borgani} \&
  {Dolag}}{{Ettori} et~al.}{2012}]{2012MNRAS.420.2058E}
{Ettori} S.,  {Rasia} E.,  {Fabjan} D.,  {Borgani} S.,    {Dolag} K.,  2012,
  \mnras, 420, 2058

\bibitem[\protect\citeauthoryear{{Fabian}}{{Fabian}}{2012}]{2012ARA&A..50..455F}
{Fabian} A.~C.,  2012, \araa, 50, 455

\bibitem[\protect\citeauthoryear{{Feldmann}}{{Feldmann}}{2015}]{2015MNRAS.449.3274F}
{Feldmann} R.,  2015, \mnras, 449, 3274

\bibitem[\protect\citeauthoryear{{Gallazzi}, {Charlot}, {Brinchmann}, {White}
  \& {Tremonti}}{{Gallazzi} et~al.}{2005}]{2005MNRAS.362...41G}
{Gallazzi} A.,  {Charlot} S.,  {Brinchmann} J.,  {White} S.~D.~M.,
  {Tremonti} C.~A.,  2005, \mnras, 362, 41

\bibitem[\protect\citeauthoryear{{Genel}, {Vogelsberger}, {Springel},
  {Sijacki}, {Nelson}, {Snyder}, {Rodriguez-Gomez}, {Torrey} \&
  {Hernquist}}{{Genel} et~al.}{2014}]{2014MNRAS.445..175G}
{Genel} S.,  {Vogelsberger} M.,  {Springel} V.,  {Sijacki} D.,  {Nelson} D.,
  {Snyder} G.,  {Rodriguez-Gomez} V.,  {Torrey} P.,    {Hernquist} L.,  2014,
  \mnras, 445, 175

\bibitem[\protect\citeauthoryear{{Giodini}, {Pierini}, {Finoguenov}, {Pratt},
  {Boehringer}, {Leauthaud}, {Guzzo}, {Aussel} \& {COSMOS
  Collaboration}}{{Giodini} et~al.}{2009}]{2009ApJ...703..982G}
{Giodini} S.,  {Pierini} D.,  {Finoguenov} A.,  {Pratt} G.~W.,  {Boehringer}
  H.,  {Leauthaud} A.,  {Guzzo} L.,  {Aussel} H.,    {COSMOS Collaboration}
  2009, \apj, 703, 982

\bibitem[\protect\citeauthoryear{{Gonzalez}, {Sivanandam}, {Zabludoff} \&
  {Zaritsky}}{{Gonzalez} et~al.}{2013}]{2013ApJ...778...14G}
{Gonzalez} A.~H.,  {Sivanandam} S.,  {Zabludoff} A.~I.,    {Zaritsky} D.,
  2013, \apj, 778, 14

\bibitem[\protect\citeauthoryear{{Governato}, {Brook}, {Mayer}, {Brooks},
  {Rhee}, {Wadsley}, {Jonsson}, {Willman}, {Stinson}, {Quinn} \&
  {Madau}}{{Governato} et~al.}{2010}]{2010Natur.463..203G}
{Governato} F.,  {Brook} C.,  {Mayer} L.,  {Brooks} A.,  {Rhee} G.,  {Wadsley}
  J.,  {Jonsson} P.,  {Willman} B.,  {Stinson} G.,  {Quinn} T.,    {Madau} P.,
  2010, \nat, 463, 203

\bibitem[\protect\citeauthoryear{{Gunn} \& {Gott} III}{{Gunn} \&
  {Gott}}{1972}]{GunnGott:72}
{Gunn} J.~E.,  {Gott} III J.~R.,  1972, \apj, 176, 1

\bibitem[\protect\citeauthoryear{{Haardt} \& {Madau}}{{Haardt} \&
  {Madau}}{1996}]{1996ApJ...461...20H}
{Haardt} F.,  {Madau} P.,  1996, \apj, 461, 20

\bibitem[\protect\citeauthoryear{{Hahn} \& {Abel}}{{Hahn} \&
  {Abel}}{2011}]{2011MNRAS.415.2101H}
{Hahn} O.,  {Abel} T.,  2011, \mnras, 415, 2101

\bibitem[\protect\citeauthoryear{{Hahn}, {Martizzi}, {Wu}, {Evrard}, {Teyssier}
  \& {Wechsler}}{{Hahn} et~al.}{2015}]{2015arXiv150904289H}
{Hahn} O.,  {Martizzi} D.,  {Wu} H.-Y.,  {Evrard} A.~E.,  {Teyssier} R.,
  {Wechsler} R.~H.,  2015, arXiv:1509.04289

\bibitem[\protect\citeauthoryear{{Hobbs}, {Power}, {Nayakshin} \&
  {King}}{{Hobbs} et~al.}{2012}]{2012MNRAS.421.3443H}
{Hobbs} A.,  {Power} C.,  {Nayakshin} S.,    {King} A.~R.,  2012, \mnras, 421,
  3443

\bibitem[\protect\citeauthoryear{{Kapferer}, {Kronberger}, {Breitschwerdt},
  {Schindler}, {van Kampen}, {Kimeswenger}, {Domainko}, {Mair} \&
  {Ruffert}}{{Kapferer} et~al.}{2009}]{2009A&A...504..719K}
{Kapferer} W.,  {Kronberger} T.,  {Breitschwerdt} D.,  {Schindler} S.,  {van
  Kampen} E.,  {Kimeswenger} S.,  {Domainko} W.,  {Mair} M.,    {Ruffert} M.,
  2009, \aap, 504, 719

\bibitem[\protect\citeauthoryear{{Kere{\v s}}, {Katz}, {Weinberg} \&
  {Dav{\'e}}}{{Kere{\v s}} et~al.}{2005}]{2005MNRAS.363....2K}
{Kere{\v s}} D.,  {Katz} N.,  {Weinberg} D.~H.,    {Dav{\'e}} R.,  2005,
  \mnras, 363, 2

\bibitem[\protect\citeauthoryear{{Kravtsov}, {Vikhlinin} \&
  {Meshscheryakov}}{{Kravtsov} et~al.}{2014}]{2014arXiv1401.7329K}
{Kravtsov} A.,  {Vikhlinin} A.,    {Meshscheryakov} A.,  2014, arXiv:1401.7329

\bibitem[\protect\citeauthoryear{{Laporte}, {White}, {Naab} \& {Gao}}{{Laporte}
  et~al.}{2013}]{2013MNRAS.435..901L}
{Laporte} C.~F.~P.,  {White} S.~D.~M.,  {Naab} T.,    {Gao} L.,  2013, \mnras,
  435, 901

\bibitem[\protect\citeauthoryear{{Lau}, {Nagai}, {Avestruz}, {Nelson} \&
  {Vikhlinin}}{{Lau} et~al.}{2015}]{2015ApJ...806...68L}
{Lau} E.~T.,  {Nagai} D.,  {Avestruz} C.,  {Nelson} K.,    {Vikhlinin} A.,
  2015, \apj, 806, 68

\bibitem[\protect\citeauthoryear{{Le Brun}, {McCarthy}, {Schaye} \&
  {Ponman}}{{Le Brun} et~al.}{2014}]{2014MNRAS.441.1270L}
{Le Brun} A.~M.~C.,  {McCarthy} I.~G.,  {Schaye} J.,    {Ponman} T.~J.,  2014,
  \mnras, 441, 1270

\bibitem[\protect\citeauthoryear{{Leccardi} \& {Molendi}}{{Leccardi} \&
  {Molendi}}{2008}]{2008A&A...487..461L}
{Leccardi} A.,  {Molendi} S.,  2008, \aap, 487, 461

\bibitem[\protect\citeauthoryear{{Lidman}, {Iacobuta}, {Bauer}, {Barrientos},
  {Cerulo}, {Couch}, {Delaye}, {Demarco}, {Ellingson} \& {Faloon}}{{Lidman}
  et~al.}{2013}]{2013MNRAS.433..825L}
{Lidman} C.,  {Iacobuta} G.,  {Bauer} A.~E.,  {Barrientos} L.~F.,  {Cerulo} P.,
   {Couch} W.~J.,  {Delaye} L.,  {Demarco} R.,  {Ellingson} E.,    {Faloon}
  A.~J. e.~a.,  2013, \mnras, 433, 825

\bibitem[\protect\citeauthoryear{{Lilly}, {Carollo}, {Pipino}, {Renzini} \&
  {Peng}}{{Lilly} et~al.}{2013}]{2013ApJ...772..119L}
{Lilly} S.~J.,  {Carollo} C.~M.,  {Pipino} A.,  {Renzini} A.,    {Peng} Y.,
  2013, \apj, 772, 119

\bibitem[\protect\citeauthoryear{{Lin}, {Brodwin}, {Gonzalez}, {Bode},
  {Eisenhardt}, {Stanford} \& {Vikhlinin}}{{Lin}
  et~al.}{2013}]{2013ApJ...771...61L}
{Lin} Y.-T.,  {Brodwin} M.,  {Gonzalez} A.~H.,  {Bode} P.,  {Eisenhardt}
  P.~R.~M.,  {Stanford} S.~A.,    {Vikhlinin} A.,  2013, \apj, 771, 61

\bibitem[\protect\citeauthoryear{{Liu}, {Mao} \& {Meng}}{{Liu}
  et~al.}{2012}]{2012MNRAS.423..422L}
{Liu} F.~S.,  {Mao} S.,    {Meng} X.~M.,  2012, \mnras, 423, 422

\bibitem[\protect\citeauthoryear{{Longobardi}, {Arnaboldi}, {Gerhard} \&
  {Hanuschik}}{{Longobardi} et~al.}{2015}]{2015A&A...579A.135L}
{Longobardi} A.,  {Arnaboldi} M.,  {Gerhard} O.,    {Hanuschik} R.,  2015,
  \aap, 579, A135

\bibitem[\protect\citeauthoryear{{Madau} \& {Rees}}{{Madau} \&
  {Rees}}{2001}]{2001ApJ...551L..27M}
{Madau} P.,  {Rees} M.~J.,  2001, \apjl, 551, L27

\bibitem[\protect\citeauthoryear{{Mantz}, {Allen}, {Morris}, {Rapetti},
  {Applegate}, {Kelly}, {von der Linden} \& {Schmidt}}{{Mantz}
  et~al.}{2014}]{2014MNRAS.440.2077M}
{Mantz} A.~B.,  {Allen} S.~W.,  {Morris} R.~G.,  {Rapetti} D.~A.,  {Applegate}
  D.~E.,  {Kelly} P.~L.,  {von der Linden} A.,    {Schmidt} R.~W.,  2014,
  \mnras, 440, 2077

\bibitem[\protect\citeauthoryear{{Martizzi}, {Jimmy}, {Teyssier} \&
  {Moore}}{{Martizzi} et~al.}{2014b}]{Martizzi2014b}
{Martizzi} D.,  {Jimmy} {Teyssier} R.,    {Moore} B.,  {2014b}, \mnras, 443,
  1500

\bibitem[\protect\citeauthoryear{{Martizzi}, {Mohammed}, {Teyssier} \&
  {Moore}}{{Martizzi} et~al.}{2014a}]{Martizzi2014a}
{Martizzi} D.,  {Mohammed} I.,  {Teyssier} R.,    {Moore} B.,  {2014a}, \mnras,
  440, 2290

\bibitem[\protect\citeauthoryear{{Martizzi}, {Teyssier} \& {Moore}}{{Martizzi}
  et~al.}{2012a}]{Martizzi2012a}
{Martizzi} D.,  {Teyssier} R.,    {Moore} B.,  {2012a}, \mnras, 420, 2859

\bibitem[\protect\citeauthoryear{{Martizzi}, {Teyssier}, {Moore} \&
  {Wentz}}{{Martizzi} et~al.}{2012b}]{Martizzi2012b}
{Martizzi} D.,  {Teyssier} R.,  {Moore} B.,    {Wentz} T.,  {2012b}, \mnras,
  422, 3081

\bibitem[\protect\citeauthoryear{{Matsushita}, {Sakuma}, {Sasaki}, {Sato} \&
  {Simionescu}}{{Matsushita} et~al.}{2013}]{2013ApJ...764..147M}
{Matsushita} K.,  {Sakuma} E.,  {Sasaki} T.,  {Sato} K.,    {Simionescu} A.,
  2013, \apj, 764, 147

\bibitem[\protect\citeauthoryear{{Mayer}, {Governato}, {Colpi}, {Moore},
  {Quinn}, {Wadsley}, {Stadel} \& {Lake}}{{Mayer}
  et~al.}{2001}]{2001ApJ...559..754M}
{Mayer} L.,  {Governato} F.,  {Colpi} M.,  {Moore} B.,  {Quinn} T.,  {Wadsley}
  J.,  {Stadel} J.,    {Lake} G.,  2001, \apj, 559, 754

\bibitem[\protect\citeauthoryear{{Mayer}, {Mastropietro}, {Wadsley}, {Stadel}
  \& {Moore}}{{Mayer} et~al.}{2006}]{2006MNRAS.369.1021M}
{Mayer} L.,  {Mastropietro} C.,  {Wadsley} J.,  {Stadel} J.,    {Moore} B.,
  2006, \mnras, 369, 1021

\bibitem[\protect\citeauthoryear{{McCarthy}, {Frenk}, {Font}, {Lacey}, {Bower},
  {Mitchell}, {Balogh} \& {Theuns}}{{McCarthy}
  et~al.}{2008}]{2008MNRAS.383..593M}
{McCarthy} I.~G.,  {Frenk} C.~S.,  {Font} A.~S.,  {Lacey} C.~G.,  {Bower}
  R.~G.,  {Mitchell} N.~L.,  {Balogh} M.~L.,    {Theuns} T.,  2008, \mnras,
  383, 593

\bibitem[\protect\citeauthoryear{{Moore}, {Katz}, {Lake}, {Dressler} \&
  {Oemler}}{{Moore} et~al.}{1996}]{1996Natur.379..613M}
{Moore} B.,  {Katz} N.,  {Lake} G.,  {Dressler} A.,    {Oemler} A.,  1996,
  \nat, 379, 613

\bibitem[\protect\citeauthoryear{{More}, {Diemer} \& {Kravtsov}}{{More}
  et~al.}{2015}]{2015arXiv150405591M}
{More} S.,  {Diemer} B.,    {Kravtsov} A.,  2015, arXiv:1504.05591

\bibitem[\protect\citeauthoryear{{Nelson}, {Gonzalez}, {Zaritsky} \&
  {Dalcanton}}{{Nelson} et~al.}{2002a}]{2002ApJ...566..103N}
{Nelson} A.~E.,  {Gonzalez} A.~H.,  {Zaritsky} D.,    {Dalcanton} J.~J.,
  {2002a}, \apj, 566, 103

\bibitem[\protect\citeauthoryear{{Nelson}, {Simard}, {Zaritsky}, {Dalcanton} \&
  {Gonzalez}}{{Nelson} et~al.}{2002b}]{2002ApJ...567..144N}
{Nelson} A.~E.,  {Simard} L.,  {Zaritsky} D.,  {Dalcanton} J.~J.,    {Gonzalez}
  A.~H.,  {2002b}, \apj, 567, 144

\bibitem[\protect\citeauthoryear{{Nelson}, {Genel}, {Vogelsberger}, {Springel},
  {Sijacki}, {Torrey} \& {Hernquist}}{{Nelson}
  et~al.}{2015}]{2015MNRAS.448...59N}
{Nelson} D.,  {Genel} S.,  {Vogelsberger} M.,  {Springel} V.,  {Sijacki} D.,
  {Torrey} P.,    {Hernquist} L.,  2015, \mnras, 448, 59

\bibitem[\protect\citeauthoryear{{Ocvirk}, {Pichon} \& {Teyssier}}{{Ocvirk}
  et~al.}{2008}]{2008MNRAS.390.1326O}
{Ocvirk} P.,  {Pichon} C.,    {Teyssier} R.,  2008, \mnras, 390, 1326

\bibitem[\protect\citeauthoryear{{Oliva-Altamirano}, {Brough}, {Jimmy} Kim-Vy,
  {Couch}, {McDermid}, {Lidman}, {von der Linden} \&
  {Sharp}}{{Oliva-Altamirano} et~al.}{2015}]{2015MNRAS.449.3347O}
{Oliva-Altamirano} P.,  {Brough} S.,  {Jimmy} Kim-Vy T.,  {Couch} W.~J.,
  {McDermid} R.~M.,  {Lidman} C.,  {von der Linden} A.,    {Sharp} R.,  2015,
  \mnras, 449, 3347

\bibitem[\protect\citeauthoryear{{Oliva-Altamirano}, {Brough}, {Lidman},
  {Couch}, {Hopkins}, {Colless}, {Taylor}, {Robotham}, {Gunawardhana} \&
  {Ponman}}{{Oliva-Altamirano} et~al.}{2014}]{2014MNRAS.440..762O}
{Oliva-Altamirano} P.,  {Brough} S.,  {Lidman} C.,  {Couch} W.~J.,  {Hopkins}
  A.~M.,  {Colless} M.,  {Taylor} E.,  {Robotham} A.~S.~G.,  {Gunawardhana}
  M.~L.~P.,    {Ponman} T. e.~a.,  2014, \mnras, 440, 762

\bibitem[\protect\citeauthoryear{{Ostriker}}{{Ostriker}}{1999}]{1999ApJ...513..252O}
{Ostriker} E.~C.,  1999, \apj, 513, 252

\bibitem[\protect\citeauthoryear{{Planck Collaboration}, {Ade}, {Aghanim},
  {Arnaud}, {Ashdown}, {Atrio-Barandela}, {Aumont}, {Baccigalupi} \&
  {Balbi}}{{Planck Collaboration} et~al.}{2013}]{2013A&A...550A.131P}
{Planck Collaboration} {Ade} P.~A.~R.,  {Aghanim} N.,  {Arnaud} M.,  {Ashdown}
  M.,  {Atrio-Barandela} F.,  {Aumont} J.,  {Baccigalupi} C.,    {Balbi} A.
  e.~a.,  2013, \aap, 550, A131

\bibitem[\protect\citeauthoryear{{Planck Collaboration}, {Ade}, {Aghanim},
  {Arnaud}, {Ashdown}, {Aumont}, {Baccigalupi}, {Banday}, {Barreiro},
  {Bartlett} \& et al.}{{Planck Collaboration}
  et~al.}{2015}]{2015arXiv150201589P}
{Planck Collaboration} {Ade} P.~A.~R.,  {Aghanim} N.,  {Arnaud} M.,  {Ashdown}
  M.,  {Aumont} J.,  {Baccigalupi} C.,  {Banday} A.~J.,  {Barreiro} R.~B.,
  {Bartlett} J.~G.,    et al. 2015, ArXiv:1502.01589

\bibitem[\protect\citeauthoryear{{Planelles}, {Borgani}, {Fabjan}, {Killedar},
  {Murante}, {Granato}, {Ragone-Figueroa} \& {Dolag}}{{Planelles}
  et~al.}{2014}]{2014MNRAS.438..195P}
{Planelles} S.,  {Borgani} S.,  {Fabjan} D.,  {Killedar} M.,  {Murante} G.,
  {Granato} G.~L.,  {Ragone-Figueroa} C.,    {Dolag} K.,  2014, \mnras, 438,
  195

\bibitem[\protect\citeauthoryear{{Puchwein}, {Sijacki} \&
  {Springel}}{{Puchwein} et~al.}{2008}]{2008ApJ...687L..53P}
{Puchwein} E.,  {Sijacki} D.,    {Springel} V.,  2008, \apjl, 687, L53

\bibitem[\protect\citeauthoryear{{Ragone-Figueroa}, {Granato}, {Murante},
  {Borgani} \& {Cui}}{{Ragone-Figueroa} et~al.}{2013}]{2013MNRAS.436.1750R}
{Ragone-Figueroa} C.,  {Granato} G.~L.,  {Murante} G.,  {Borgani} S.,    {Cui}
  W.,  2013, \mnras, 436, 1750

\bibitem[\protect\citeauthoryear{{Rasia}, {Borgani}, {Murante}, {Planelles},
  {Beck}, {Biffi}, {Ragone-Figueroa}, {Granato}, {Steinborn} \&
  {Dolag}}{{Rasia} et~al.}{2015}]{2015arXiv150904247R}
{Rasia} E.,  {Borgani} S.,  {Murante} G.,  {Planelles} S.,  {Beck} A.~M.,
  {Biffi} V.,  {Ragone-Figueroa} C.,  {Granato} G.~L.,  {Steinborn} L.~K.,
  {Dolag} K.,  2015, arXiv:1509.04247

\bibitem[\protect\citeauthoryear{{Renzini} \& {Andreon}}{{Renzini} \&
  {Andreon}}{2014}]{2014MNRAS.444.3581R}
{Renzini} A.,  {Andreon} S.,  2014, \mnras, 444, 3581

\bibitem[\protect\citeauthoryear{{Romeo}, {Sommer-Larsen}, {Portinari} \&
  {Antonuccio-Delogu}}{{Romeo} et~al.}{2006}]{2006MNRAS.371..548R}
{Romeo} A.~D.,  {Sommer-Larsen} J.,  {Portinari} L.,    {Antonuccio-Delogu} V.,
   2006, \mnras, 371, 548

\bibitem[\protect\citeauthoryear{{Ruszkowski} \& {Springel}}{{Ruszkowski} \&
  {Springel}}{2009}]{2009ApJ...696.1094R}
{Ruszkowski} M.,  {Springel} V.,  2009, \apj, 696, 1094

\bibitem[\protect\citeauthoryear{{Schaye}, {Crain}, {Bower}, {Furlong},
  {Schaller}, {Theuns}, {Dalla Vecchia}, {Frenk}, {McCarthy}, {Helly},
  {Jenkins}, {Rosas-Guevara}, {White}, {Baes}, {Booth}, {Camps}, {Navarro},
  {Qu} \& {Rahmati}}{{Schaye} et~al.}{2015}]{2015MNRAS.446..521S}
{Schaye} J.,  {Crain} R.~A.,  {Bower} R.~G.,  {Furlong} M.,  {Schaller} M.,
  {Theuns} T.,  {Dalla Vecchia} C.,  {Frenk} C.~S.,  {McCarthy} I.~G.,  {Helly}
  J.~C.,  {Jenkins} A.,  {Rosas-Guevara} Y.~M.,  {White} S.~D.~M.,  {Baes} M.,
  {Booth} C.~M.,  {Camps} P.,  {Navarro} J.~F.,  {Qu} Y.,    {Rahmati} A.
  e.~a.,  2015, \mnras, 446, 521

\bibitem[\protect\citeauthoryear{{Segers}, {Crain}, {Schaye}, {Bower},
  {Furlong}, {Schaller} \& {Theuns}}{{Segers}
  et~al.}{2015}]{2015arXiv150708281S}
{Segers} M.~C.,  {Crain} R.~A.,  {Schaye} J.,  {Bower} R.~G.,  {Furlong} M.,
  {Schaller} M.,    {Theuns} T.,  2015, arXiv:1507.08281

\bibitem[\protect\citeauthoryear{{Sembolini}, {Yepes}, {Pearce}, {Knebe},
  {Kay}, {Power}, {Cui} \& {Beck}}{{Sembolini}
  et~al.}{2015}]{2015arXiv150306065S}
{Sembolini} F.,  {Yepes} G.,  {Pearce} F.~R.,  {Knebe} A.,  {Kay} S.~T.,
  {Power} C.,  {Cui} W.,    {Beck} A.~M. e.~a.,  2015, arXiv:1503.06065

\bibitem[\protect\citeauthoryear{{Short}, {Thomas} \& {Young}}{{Short}
  et~al.}{2013}]{2013MNRAS.428.1225S}
{Short} C.~J.,  {Thomas} P.~A.,    {Young} O.~E.,  2013, \mnras, 428, 1225

\bibitem[\protect\citeauthoryear{{Sijacki} \& {Springel}}{{Sijacki} \&
  {Springel}}{2006}]{Sijacki:2006}
{Sijacki} D.,  {Springel} V.,  2006, \mnras, 366, 397

\bibitem[\protect\citeauthoryear{{Sijacki}, {Springel}, {Di Matteo} \&
  {Hernquist}}{{Sijacki} et~al.}{2007}]{2007MNRAS.380..877S}
{Sijacki} D.,  {Springel} V.,  {Di Matteo} T.,    {Hernquist} L.,  2007,
  \mnras, 380, 877

\bibitem[\protect\citeauthoryear{{Skory}, {Hallman}, {Burns}, {Skillman},
  {O'Shea} \& {Smith}}{{Skory} et~al.}{2013}]{2013ApJ...763...38S}
{Skory} S.,  {Hallman} E.,  {Burns} J.~O.,  {Skillman} S.~W.,  {O'Shea} B.~W.,
    {Smith} B.~D.,  2013, \apj, 763, 38

\bibitem[\protect\citeauthoryear{{Springel}}{{Springel}}{2005}]{2005MNRAS.364.1105S}
{Springel} V.,  2005, \mnras, 364, 1105

\bibitem[\protect\citeauthoryear{{Springel}}{{Springel}}{2010}]{2010MNRAS.401..791S}
{Springel} V.,  2010, \mnras, 401, 791

\bibitem[\protect\citeauthoryear{{Springel}, {Di Matteo} \&
  {Hernquist}}{{Springel} et~al.}{2005}]{2005MNRAS.361..776S}
{Springel} V.,  {Di Matteo} T.,    {Hernquist} L.,  2005, \mnras, 361, 776

\bibitem[\protect\citeauthoryear{{Sutherland} \& {Dopita}}{{Sutherland} \&
  {Dopita}}{1993}]{1993ApJS...88..253S}
{Sutherland} R.~S.,  {Dopita} M.~A.,  1993, \apjs, 88, 253

\bibitem[\protect\citeauthoryear{{Teyssier}}{{Teyssier}}{2002}]{2002A&A...385..337T}
{Teyssier} R.,  2002, \aap, 385, 337

\bibitem[\protect\citeauthoryear{{Teyssier}, {Moore}, {Martizzi}, {Dubois} \&
  {Mayer}}{{Teyssier} et~al.}{2011}]{Teyssier:2011}
{Teyssier} R.,  {Moore} B.,  {Martizzi} D.,  {Dubois} Y.,    {Mayer} L.,  2011,
  \mnras, 414, 195

\bibitem[\protect\citeauthoryear{{Teyssier}, {Pontzen}, {Dubois} \&
  {Read}}{{Teyssier} et~al.}{2013}]{2013MNRAS.429.3068T}
{Teyssier} R.,  {Pontzen} A.,  {Dubois} Y.,    {Read} J.~I.,  2013, \mnras,
  429, 3068

\bibitem[\protect\citeauthoryear{{Tonini}, {Bernyk}, {Croton}, {Maraston} \&
  {Thomas}}{{Tonini} et~al.}{2012}]{2012ApJ...759...43T}
{Tonini} C.,  {Bernyk} M.,  {Croton} D.,  {Maraston} C.,    {Thomas} D.,  2012,
  \apj, 759, 43

\bibitem[\protect\citeauthoryear{{Tornatore}, {Borgani}, {Dolag} \&
  {Matteucci}}{{Tornatore} et~al.}{2007}]{2007MNRAS.382.1050T}
{Tornatore} L.,  {Borgani} S.,  {Dolag} K.,    {Matteucci} F.,  2007, \mnras,
  382, 1050

\bibitem[\protect\citeauthoryear{{Trayford}, {Theuns}, {Bower}, {Schaye},
  {Furlong}, {Schaller}, {Frenk}, {Crain}, {Dalla Vecchia} \&
  {McCarthy}}{{Trayford} et~al.}{2015}]{2015MNRAS.452.2879T}
{Trayford} J.~W.,  {Theuns} T.,  {Bower} R.~G.,  {Schaye} J.,  {Furlong} M.,
  {Schaller} M.,  {Frenk} C.~S.,  {Crain} R.~A.,  {Dalla Vecchia} C.,
  {McCarthy} I.~G.,  2015, \mnras, 452, 2879

\bibitem[\protect\citeauthoryear{{Trujillo}, {F{\"o}rster Schreiber},
  {Rudnick}, {Barden}, {Franx}, {Rix}, {Caldwell} \& {McIntosh}}{{Trujillo}
  et~al.}{2006}]{2006ApJ...650...18T}
{Trujillo} I.,  {F{\"o}rster Schreiber} N.~M.,  {Rudnick} G.,  {Barden} M.,
  {Franx} M.,  {Rix} H.-W.,  {Caldwell} J.~A.~R.,    {McIntosh} D.~H. e.~a.,
  2006, \apj, 650, 18

\bibitem[\protect\citeauthoryear{{Valdarnini}}{{Valdarnini}}{2003}]{2003MNRAS.339.1117V}
{Valdarnini} R.,  2003, \mnras, 339, 1117

\bibitem[\protect\citeauthoryear{{van de Voort}, {Schaye}, {Booth}, {Haas} \&
  {Dalla Vecchia}}{{van de Voort} et~al.}{2011}]{2011MNRAS.414.2458V}
{van de Voort} F.,  {Schaye} J.,  {Booth} C.~M.,  {Haas} M.~R.,    {Dalla
  Vecchia} C.,  2011, \mnras, 414, 2458

\bibitem[\protect\citeauthoryear{{van den Bosch}, {Aquino}, {Yang}, {Mo},
  {Pasquali}, {McIntosh}, {Weinmann} \& {Kang}}{{van den Bosch}
  et~al.}{2008}]{2008MNRAS.387...79V}
{van den Bosch} F.~C.,  {Aquino} D.,  {Yang} X.,  {Mo} H.~J.,  {Pasquali} A.,
  {McIntosh} D.~H.,  {Weinmann} S.~M.,    {Kang} X.,  2008, \mnras, 387, 79

\bibitem[\protect\citeauthoryear{{van der Wel}, {Franx}, {van Dokkum},
  {Skelton}, {Momcheva}, {Whitaker}, {Brammer}, {Bell}, {Rix} \& {Wuyts}}{{van
  der Wel} et~al.}{2014}]{2014ApJ...788...28V}
{van der Wel} A.,  {Franx} M.,  {van Dokkum} P.~G.,  {Skelton} R.~E.,
  {Momcheva} I.~G.,  {Whitaker} K.~E.,  {Brammer} G.~B.,  {Bell} E.~F.,  {Rix}
  H.-W.,    {Wuyts} S. e.~a.,  2014, \apj, 788, 28

\bibitem[\protect\citeauthoryear{{van der Wel}, {Holden}, {Franx},
  {Illingworth}, {Postman}, {Kelson}, {Labb{\'e}}, {Wuyts}, {Blakeslee} \&
  {Ford}}{{van der Wel} et~al.}{2007}]{2007ApJ...670..206V}
{van der Wel} A.,  {Holden} B.~P.,  {Franx} M.,  {Illingworth} G.~D.,
  {Postman} M.~P.,  {Kelson} D.~D.,  {Labb{\'e}} I.,  {Wuyts} S.,  {Blakeslee}
  J.~P.,    {Ford} H.~C.,  2007, \apj, 670, 206

\bibitem[\protect\citeauthoryear{{Vogelsberger}, {Genel}, {Springel}, {Torrey},
  {Sijacki}, {Xu}, {Snyder}, {Nelson} \& {Hernquist}}{{Vogelsberger}
  et~al.}{2014}]{2014MNRAS.444.1518V}
{Vogelsberger} M.,  {Genel} S.,  {Springel} V.,  {Torrey} P.,  {Sijacki} D.,
  {Xu} D.,  {Snyder} G.,  {Nelson} D.,    {Hernquist} L.,  2014, \mnras, 444,
  1518

\bibitem[\protect\citeauthoryear{{Wetzel} \& {Nagai}}{{Wetzel} \&
  {Nagai}}{2014}]{2014arXiv1412.0662W}
{Wetzel} A.~R.,  {Nagai} D.,  2014, arXiv:1412.0662

\bibitem[\protect\citeauthoryear{{Whiley}, {Arag{\'o}n-Salamanca}, {De Lucia},
  {von der Linden}, {Bamford}, {Best}, {Bremer}, {Jablonka}, {Johnson},
  {Milvang-Jensen}, {Noll}, {Poggianti}, {Rudnick}, {Saglia}, {White} \&
  {Zaritsky}}{{Whiley} et~al.}{2008}]{2008MNRAS.387.1253W}
{Whiley} I.~M.,  {Arag{\'o}n-Salamanca} A.,  {De Lucia} G.,  {von der Linden}
  A.,  {Bamford} S.~P.,  {Best} P.,  {Bremer} M.~N.,  {Jablonka} P.,  {Johnson}
  O.,  {Milvang-Jensen} B.,  {Noll} S.,  {Poggianti} B.~M.,  {Rudnick} G.,
  {Saglia} R.,  {White} S.,    {Zaritsky} D.,  2008, \mnras, 387, 1253

\bibitem[\protect\citeauthoryear{{Wiersma}, {Schaye} \& {Theuns}}{{Wiersma}
  et~al.}{2011}]{2011MNRAS.415..353W}
{Wiersma} R.~P.~C.,  {Schaye} J.,    {Theuns} T.,  2011, \mnras, 415, 353

\bibitem[\protect\citeauthoryear{{Woosley} \& {Weaver}}{{Woosley} \&
  {Weaver}}{1995}]{1995ApJS..101..181W}
{Woosley} S.~E.,  {Weaver} T.~A.,  1995, \apjs, 101, 181

\bibitem[\protect\citeauthoryear{{Wu}, {Evrard}, {Hahn}, {Martizzi}, {Teyssier}
  \& {Wechsler}}{{Wu} et~al.}{2015}]{2015MNRAS.452.1982W}
{Wu} H.-Y.,  {Evrard} A.~E.,  {Hahn} O.,  {Martizzi} D.,  {Teyssier} R.,
  {Wechsler} R.~H.,  2015, \mnras, 452, 1982

\bibitem[\protect\citeauthoryear{{Wu}, {Hahn}, {Wechsler}, {Behroozi} \&
  {Mao}}{{Wu} et~al.}{2013b}]{Wu2013b}
{Wu} H.-Y.,  {Hahn} O.,  {Wechsler} R.~H.,  {Behroozi} P.~S.,    {Mao} Y.-Y.,
  {2013b}, \apj, 767, 23

\bibitem[\protect\citeauthoryear{{Wu}, {Hahn}, {Wechsler}, {Mao} \&
  {Behroozi}}{{Wu} et~al.}{2013a}]{Wu2013a}
{Wu} H.-Y.,  {Hahn} O.,  {Wechsler} R.~H.,  {Mao} Y.-Y.,    {Behroozi} P.~S.,
  {2013a}, \apj, 763, 70

\end{thebibliography}

\appendix
\section{Metal yield from unresolved stellar populations}\label{appendix:A}
In this Appendix we try to quantify the amount of metals that are not produced in the simulations because limited resolution does not allow to resolve the 
formation of the whole galaxy population. In an ideal simulation with infinite resolution the stellar mass within the virial radius would be:
\begin{equation}
M_{*}(<R_{\rm vir}) = \int_{M_{\rm min}}^{M_{\rm max}} dM_{*} \phi(M_{*}) M_{*} + M_{\rm ICL},
\end{equation}
where $\phi(M_{*})$ is the real SMF, $M_{ICL}$ is the mass in the intracluster light (ICL) and $M_{\rm min}=10^{8}$ M$_{\odot}$ is a minimum mass threshold 
and $M_{\rm max}=1.23\times 10^{11}$ M$_{\odot}$ is the characteristic mass scale of the Schechter function. 
In a simulation with finite resolution, the stellar mass within the virial radius would be:
\begin{equation}
M_{\rm *,res}(<R_{\rm vir}) = \int_{M_{\rm min}}^{M_{\rm max}} dM_{*} \phi_{\rm res}(M_{*}) M_{*} + M_{\rm ICL,res},
\end{equation}
where $\phi_{\rm res}(M_{*})$ is the SMF in the simulation and $M_{\rm ICL,res}$ is the mass of the ICL resolved in the simulations. Under the assumption that 
$M_{\rm ICL,res}\approx M_{\rm ICL}$, the mass difference between the ideal case and the normal simulation is:
\begin{equation}
 \Delta M_{*}=M_{*}-M_{\rm *,res}=\int_{M_{\rm min}}^{M_{\rm max}} dM_{*} M_{*} [\phi(M_{*})-\phi_{\rm res}(M_{*})].
\end{equation}
Now we proceed to compute the total mass in metals injected by supernovae. In our simulations the mass ejected by supernovae is a fraction $\eta=0.1$ of the 
initial gaseous mass that is converted into stars; a fraction $y=0.1$ of this mass is made of metals. If we neglect the fact that some metals will end up in 
newly formed stars, then the total mass in metals expected within the virial radius in a normal simulations is just given by
\begin{equation}
 M_{\rm Z,res}\approx\frac{y\eta}{1-\eta}M_{\rm *,res}.
\end{equation}
In the ideal case with infinite resolution we would get a metal mass equal to 
\begin{equation}
 M_{\rm Z}\approx\frac{y\eta}{1-\eta}M_{\rm *}=\frac{y\eta}{1-\eta}(M_{\rm *,res}+\Delta M_{*}).
\end{equation}
The final result is that the unresolved metal mass in the halo can be written as: 
\begin{equation}\label{miss_metals}
 \Delta M_{\rm Z}\approx\frac{y\eta}{1-\eta}\Delta M_{\rm *}
\end{equation}
A very important caveat is that eq.~(\ref{miss_metals}) deals only with integrated quantities, i.e. it applies to the halo as a whole and does not 
include any information about the spatial distribution of metals, which can be a very relevant issue. The unresolved metallicity can be estimated by dividing 
the unresolved metal mass by the gaseous mass:
\begin{equation}
 {\rm Z}_{\rm ur}(<R_{\rm vir})\approx\frac{\Delta M_{\rm Z}(<R_{\rm vir})}{M_{\rm gas}(<R_{\rm vir})}.
\end{equation}
The values of the unresolved metallicities in each halo available at redshift $z=0$ are summarised in Table\ref{tab:urmetals}.

\begin{table*}
\begin{center}
{\bfseries Unresolved metals at redshift $z=0$}
\end{center}
\begin{center}
\begin{tabular}{lcccc}
\hline
 Halo & Resolution & $M_{\rm gas}(<R_{\rm vir})$ [M$_{\odot}$]& $\Delta M_{\rm Z}(<R_{\rm vir})$ [M$_{\odot}$] & ${\rm Z}_{\rm ur}(<R_{\rm vir})$ [Z$_{\odot}$]\\
\hline
211 & R4K & $1.65\times 10^{14}$ & $1.96\times 10^{10}$ & $5.93\times 10^{-3}$ \\
337 & R4K & $1.80\times 10^{14}$ & $3.54\times 10^{10}$ & $9.82\times 10^{-3}$ \\ 
348 & R4K & $2.01\times 10^{14}$ & $4.09\times 10^{10}$ & $1.03\times 10^{-2}$ \\ 
361 & R4K & $1.85\times 10^{14}$ & $2.96\times 10^{10}$ & $8.02\times 10^{-3}$ \\
377 & R4K & $1.73\times 10^{14}$ & $2.45\times 10^{10}$ & $7.09\times 10^{-3}$ \\
448 & R4K & $1.71\times 10^{14}$ & $1.99\times 10^{10}$ & $5.79\times 10^{-3}$ \\
474 & R4K & $4.49\times 10^{14}$ & $2.33\times 10^{11}$ & $2.59\times 10^{-2}$ \\
545 & R4K & $1.64\times 10^{14}$ & $2.71\times 10^{10}$ & $8.23\times 10^{-3}$ \\
572 & R4K & $1.74\times 10^{14}$ & $3.32\times 10^{10}$ & $9.54\times 10^{-3}$ \\
653 & R4K & $1.35\times 10^{14}$ & $7.47\times 10^{9}$  & $2.77\times 10^{-3}$ \\
653 & R8K & $1.33\times 10^{14}$ & $6.90\times 10^{8}$  & $2.59\times 10^{-4}$ \\
\hline
\end{tabular}
\end{center}
\caption{ The table contains the value for several quantities related to metal yield from the unresolved population of satellite galaxies within the virial 
radius. The values are given for all the simulations available at redshift $z=0$. Col. 1: halo ID number. Col. 2: resolution. Col. 3: gaseous mass within 
the virial radius. Col. 4: unresolved metal mass within the virial radius. Col. 5: unresolved metallicity within the virial radius. }\label{tab:urmetals}
\end{table*}

Given the values in Table~\ref{tab:urmetals} it is possible to come up with a simple model to correct for the missing metal yield in simulations with unresolved 
stellar populations. The mean unresolved metallicity (with standard deviation) among all the 4K runs is 
\begin{equation}
<{\rm Z}_{\rm ur}>_{\rm R4K}=(8\pm6)\times 10^{-3} \hbox{Z}_{\odot}.
\end{equation}
Considering the way the unresolved metallicity scales with resolution in halo 653, we propose a formula for the missing metallicity as a function of resolution:
\begin{equation}
 {\rm Z}_{\rm ur} (M_{\rm *,res})= 8\times 10^{-3}  \hbox{Z}_{\odot} +2.5\times 10^{-3} \hbox{Z}_{\odot} \times (\log_{10}M_{\rm *,res}-10.5),
\end{equation}
where $M_{\rm *,res}$ is the minimum resolved galaxy stellar mass and corresponds to $\sim 500$ particles in our simulations. 

\section{Effect of the metal yield from star formation}\label{appendix:b}
\begin{figure}
\begin{center}
\includegraphics[width=0.49\textwidth]{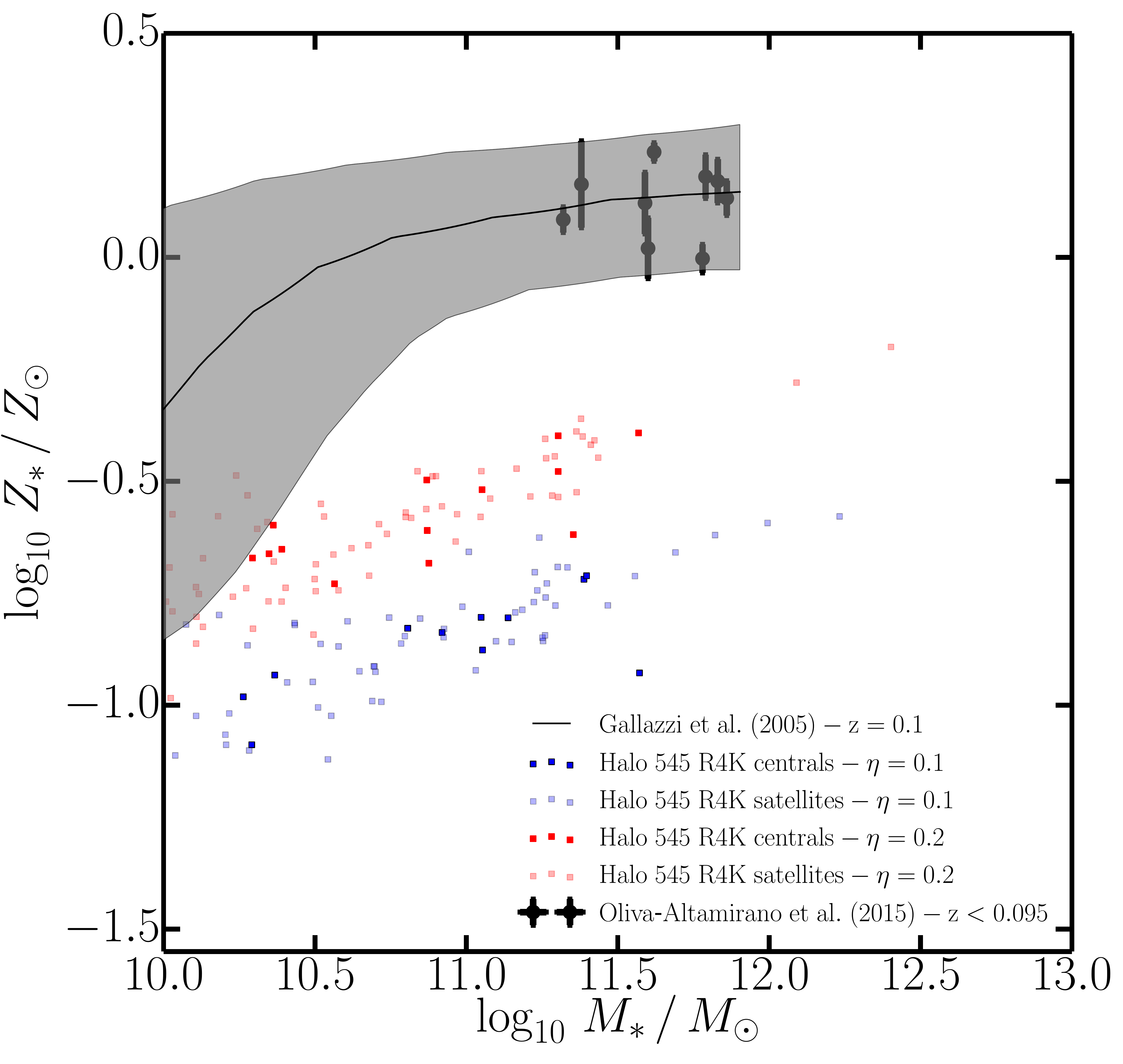}
\end{center}
\caption{\label{fig:mstar_zstar_yield_test} Stellar mass-metallicity relation at low redshift for the galaxies in halo 545 R4K with $y=0.1$ (blue) and for the galaxies in halo 545 R4K with $y=0.2$(red). Centrals are represented by full colour squares, satellites by semi-transparent squares. The simulations are 
compared to the BCG data from Oliva-Altamirano et al. (2015; black points with error bars) and to the global stellar mass-metallicity relation determined 
by Gallazzi et al. (2005; black line with shaded area representing the 1-$\sigma$ scatter). }
\end{figure}

In our fiducial {\sc Rhapsody-G} simulations we always assumed a stellar yield $y=0.1$. This value for the metal yield has been chosen to match, on average, the metal enrichment from massive stars as computed by \cite{1995ApJS..101..181W}. However, the metal yield of a given star depends on its composition and evolution and this leaves some freedom to choose the average stellar yield of a stellar population. For example, \cite{2013ApJ...772..119L} assume a metal yield as high as four times the value we adopted in {\sc Rhapsody-G}. We performed a test of the impact of $y$ on our results by re-simulating halo 545 with metal yield $y=0.2$, i.e. twice as large as in the original run. The stellar mass-metallicity relation of this re-simulation is compared to that of the original one in Figure~\ref{fig:mstar_zstar_yield_test}. The stellar mass in the simulated galaxies does not change much between the two runs, but with twice the yield the stellar metallicity doubles too. Combining this result with those in the main paper, we conclude that a combination of high resolution and higher metal yield might alleviate (and possibly get rid of) the discrepancy between observed metallicities and the simulations.

\label{lastpage}

\end{document}